\documentclass[twoside]{article}

\usepackage{times}\usepackage[margin=1in]{geometry}
\usepackage[round]{natbib}

\usepackage{enumitem}

\usepackage[utf8]{inputenc} 
\usepackage[T1]{fontenc}    
\usepackage{hyperref}       
\usepackage{url}            
\usepackage{booktabs}       
\usepackage{amsfonts}       
\usepackage{nicefrac}       
\usepackage{microtype}      
\usepackage{xcolor}         
\usepackage{graphicx}
\usepackage{bm}
\usepackage{amsmath,amssymb}
\usepackage{algorithm}
\usepackage{algpseudocode}
\usepackage{soul}

\usepackage[utf8]{inputenc}

\usepackage{amsmath}
\usepackage{amsfonts}
\usepackage{amsthm}
\usepackage{booktabs}       
\usepackage{color}

\usepackage{xspace}         
\usepackage[conf={no link to proof, restate}]{proof-at-the-end}

\newcommand{\blue}[1]{{#1}}
\newcommand{\red}[1]{{#1}}
\newcommand{\eat}[1]{}

\newlength{\figwidth}
\usepackage{subcaption}

\newtheorem*{thm*}{Theorem}

\newtheorem*{lemma*}{Lemma}

\newtheorem*{fact*}{Fact}

\theoremstyle{definition}
\newtheorem{definition}{Definition}

\DeclareMathOperator{\AUC}{AUC}
\newcommand{\Var}{\mathsf{Var}}
\newcommand{\E}{\mathsf{E}}
\newcommand{\clients}{\ensuremath{M}}
\newcommand{\para}[1]{\noindent\textbf{#1}}
\newcommand{\cdp}{DistDP\xspace}
\newcommand{\ldp}{LocalDP\xspace}
\newcommand{\fed}{Federated\xspace}
\newcommand{\pred}{\ensuremath{\operatorname{pred}}}

\begin{document}



\title{Federated Calibration and Evaluation of Binary Classifiers}\author{Graham Cormode\thanks{Meta AI, \texttt{gcormode@meta.com}} \and Igor Markov\thanks{Meta AI, \texttt{imarkov@meta.com}}}\date{}\maketitle

\begin{abstract}
    We address two major obstacles to practical use of supervised classifiers on distributed private data.
    Whether a classifier was trained by a federation of cooperating clients or
    trained centrally out of distribution,
    (1) the output scores must be calibrated, and (2)  performance metrics must be evaluated --- all without assembling labels in one place. In particular, we show how to perform calibration and compute precision, recall, accuracy and ROC-AUC
    in the federated setting under three privacy models ($i$) secure aggregation, ($ii$) \red{distributed} differential privacy, ($iii$) local differential privacy. Our theorems and experiments clarify tradeoffs between privacy, accuracy, and data efficiency. 
    They also help decide whether a given application has sufficient data to support federated calibration and evaluation.
\end{abstract}

\section{Introduction}
\allowdisplaybreaks
Modern machine learning (ML) draws insights from large amounts of data, e.g., by training 
\blue{prediction models on collected labels. Traditional ML \eat{techniques and their software implementations}workflows assemble all data} in one place, but \eat{today much of}
human-generated data --- the content viewed online and reactions to this content, geographic locations, the text typed, sound and images recorded by portable devices, interactions with friends, interactions with online ads, online purchases, etc --- is subject to privacy constraints and cannot be shared easily~\citep{Cohen:Nissim:20}. This raises a major challenge: extending traditional ML techniques to accommodate a federation of cooperating distributed clients with individually private data.
In a general framework to address this challenge,
instead of collecting the data for centralized processing, clients evaluate a candidate model on their labeled data, and send updates to a server for aggregation. For example, {\em federated training}
performs learning locally on the data in parallel
in a privacy-respecting manner: the updates are gradients that are summed by the server and then used to revise the model~\citep{fedavg}. 

\para{Federated learning} \citep{flsurvey} can offer privacy guarantees to clients. 
In a baseline level of disclosure limitation, the clients never share raw data but only send out model updates. Formal privacy guarantees are obtained via careful aggregation and by adding noise to updates~\citep{fldp}. Given the prominence of federated learning, different aspects of the training process (usually for neural networks) have received much attention: aiming to optimize training speed, reduce communication and tighten privacy guarantees. However, deploying trained models usually requires to ($i$) evaluate and track their performance on distributed (private) data, ($ii$) select the best model from available alternatives, and ($iii$) calibrate a given model to frequent snapshots of non-stationary data (common in industry applications).
{\em Strong privacy guarantees for client data during the model training must be matched by similar protections for the entire ML pipeline.} Otherwise, divulging private information of clients during product use would negate earlier protections.\eat{calibration and evaluation metrics may leak information about client data (e.g.,} E.g., knowing if some label agrees with model prediction can effectively reveal the private label. 
\red{Concretely, \citet{Matthews:Harel:13} showed that disclosure of an ROC curve allows some recovery of the sensitive input data.}
In this paper, we develop algorithms for ($i$) federated evaluation of classifier-quality metrics with privacy guarantees and ($ii$) performing classifier calibration. See a summary of our results in Section \ref{sec:summary} and Table \ref{tab:results}.

\para{Federated calibration and evaluation} of classifiers are two fundamental tasks that arise 
regardless of federated learning and deep learning~\citep{RN2020}.
They are important whenever a classifier is used ``in the wild'' by distributed clients on data that\eat{could deviate from the training and test data, or where such data} cannot be collected centrally due to privacy concerns. 
\blue{Common scenarios that arise in practical\eat{distributed classifier} deployments include the following.}%
\begin{itemize}
\vspace{-2mm}
\itemsep-0mm 
    \item 
    A heuristic rule-based model is proposed as a baseline for a task but must be evaluated to understand whether a more complex ML-based solution is even needed;
    \item
    A model pre-trained via transfer learning for addressing multiple tasks (e.g., BERT~\citep{DevlinCLT19}) is to be evaluated for its performance on a particular task;%
    \item
    A new model has been trained with federated learning on fresh data, and needs to be compared with earlier models on distributed test data before launching;
    \item
    A model has been deployed to production, but its predictions must be continuously evaluated against user behavior (e.g., click-through rate) to determine when model retraining is needed.
\end{itemize}
%
%
Federated calibration of a classifier score function 
remaps raw score values to probabilities, so that examples assigned probability $p$ are positive (approximately) a $p$ proportion of the time. Since classifier decisions are routinely made by comparing the evaluated score to a threshold, calibration ensures the validity of the threshold (especially for nonstationary data) and the transparency/explainability of the classification procedure.
As noted by \citet{GuoPSW17}, ``modern neural networks are not well calibrated'', prompting recent study of calibration techniques~\citep{MDRHZHTL21}. 
Calibration (if done well) does not affect precision-recall tradeoffs.

Federated evaluation of standard binary-classifier metrics is
the second challenge we address:
precision, recall, accuracy and ROC-AUC~\citep{RN2020}. 
Accuracy measures the fraction of predictions that are correct, while recall and precision focus on only examples from a particular class, giving the fraction of these examples that are predicted correctly, 
and the fraction of examples labeled as being in the class that are correct, respectively. 
ROC-AUC is defined in terms of the area under the curve of a plot of the tradeoff between false positive ratio and true positive ratio. 
These simple metrics are easy to compute 
when the test data is \red{held centrally}.
It is more challenging to compute them in a distributed setting, while providing formal guarantees of privacy and accuracy. 
For privacy, we leverage secure aggregation and/or differential privacy (Section~\ref{sec:privacymodels}). 
For accuracy, we seek error bounds when estimating metrics, so as to facilitate practical use. 
We present bounds as a function of the number of participating clients ($\clients$) and privacy parameters ($\epsilon$). 
For the evaluation metrics we studied, errors decrease as low polynomials of $M$ --- good news for medium-to-large deployments.

\noindent
The {\bf federated computational techniques} we use are instrumental for both challenges addressed in this paper. 
These techniques compile statistical descriptions of the classifier score function as evaluated on the examples held by distributed clients. 
Then we directly estimate classifier quality, ROC AUC, and the calibration mapping. 
The description we need is essentially a histogram of the score distribution, whose buckets divide up the examples evenly.
Estimation error drops as this description gets refined: for a $B$-bucket histogram, estimation error due to the histogram scales as $1/B$ or even $1/B^2$ in some cases. 
Hence, $B \leq 100$ leads to very accurate results. 
%
Fortunately, the histogram-based approach is compatible with various privacy models that provide strong guarantees under different scenarios by adding noise to data and combining it.
The novelty in our work is in showing the reductions of these important problems to histogram computations, and in analyzing the resulting accuracy bounds.  
\eat{in different ways.} 
The privacy noise adds a term based on the number of clients $\clients$: $\clients^{-1/4}$ (worst case) or $\clients^{-1}$ (the best case).
Hence, we obtain good accuracy (under privacy) with upwards of 10,000 participating clients.

\para{The three models of privacy} considered in this work are:
(1)    Federated privacy via Secure Aggregation (\fed for short), where 
    the protocol reveals the true output of the computed aggregate (without noise addition)~\citep{secagg}.  The secrecy of the client's inputs is achieved by using Secure Aggregation to gather their values, only revealing the aggregate (typically, the sum); 
 (2) \red{Distributed Differential Privacy (\cdp), where 
 each client introduces a small amount of noise, so that the sum of all these noise fragments is equivalent in distribution to a central noise distribution such as Laplace or Gaussian, leading to a 
 guarantee of differential privacy~\citep{Dwork:Roth:14}.  Using Secure Aggregation then ensures that only the sum of inputs with privacy noise is revealed}; and
(3)    Local Differential Privacy (\ldp), where 
    each client adds sufficient noise to their report to ensure differential privacy on their message, so that Secure Aggregation is not needed~\citep{survey1}. 
These models imply different error bounds as we trade accuracy for the level of privacy and trust needed. 

\para{Our contributions} offer algorithmic techniques and error bounds for federated calibration of classifier scores and key classifier quality metrics. 
These honor three different privacy models (above). 
For local and \red{distributed} differential privacy we use the standard $\epsilon$ privacy parameter (see Section \ref{sec:privacymodels}), along with the $B$ and $M$ parameters (above) and the 
$h$ parameter explained in Section \ref{sec:scorehist}.
Our theoretical bounds are summarized in a $4\times3$ matrix seen in Table~\ref{tab:results}, and explained in greater detail in Section~\ref{sec:summary}.
\blue{Several key questions we address have largely eluded prior studies in the federated setting, and
even modest asymptotic improvements over prior results are significant in practice due to large values of $M$.
To better understand the volume of data needed to achieve acceptable accuracy bounds, 
we perform experimental studies}.  Compared to a recent result on ROC AUC estimation under \ldp~\citep{BellBGK20}, we \eat{slightly}\blue{asymptotically} improve bounds under \red{less restrictive} assumptions. 
\red{Most recently, heuristics have been proposed for AUC estimation based on local noise addition~\cite{Sunetal1, Sunetal2}.  
These do not provide any accuracy guarantees, and we see that our approach provides better results in our experimental study. 
Similar questions have also been studied in the centralized model of DP~\citep{StoddardCM14}, but these too lack the accuracy guarantees we can provide.}


\para{The paper outline} is as follows. Section~\ref{sec:prelims} reviews technical background on 
classifier calibration and evaluation, as well as privacy models. It also states the technical assumptions we use in proofs. Section~\ref{sec:scorehist} develops technical machinery for federated aggregation using score histograms. 
These histograms are needed ($a$) in Section~\ref{sec:pramain} for federated evaluation of precision, recall and accuracy of binary classifiers, ($b$) in Section~\ref{sec:rocauc-bounds} for ROC AUC, and ($c$) in Section~\ref{sec:calibration} for calibration. Empirical evaluation of our techniques is reported in Section~\ref{sec:experiments}, and conclusions are drawn in Section~\ref{sec:conclusions}.
\section{Background, Notation and Assumptions}
\label{sec:prelims}
\blue{Supervised binary classification supports many practical applications, and its theoretical setting is conducive for formal analyses. 
It also helps address multiclass classifications and subset selection (via indicator classifiers), while lightweight ranking is routinely implemented by sorting binary classifier scores trained to predict which items are more likely to be selected. 
Standard  classifier metrics are often computed approximately in practice (e.g., by monte carlo sampling), but this becomes more challenging in the federated setting.}

\subsection{Classifier Calibration and Evaluation Metrics}
\label{sec:defs}
Given a trained score function $s(\cdot)$ which takes in examples $x$ and outputs a score $s(x) \in [0,1]$, we define a binary classifier based on a threshold $T$, via
\begin{align*}
    \pred(x) = 
    \begin{cases} 0 & \text{if } s(x) \le T \\
    1 & \text{if } s(x) > T
    \end{cases}. \newline
\end{align*}
\eat{Varying $T$ from 0 to 1 trades off the false positives and true positives.}
At $T\!=\!0$, the {\em false positive ratio} (FPR) and {\em true positive ratio} (TPR) are 1, and drop to 0 as $T\!\rightarrow\!1$. 

\para{Well-behaved score distributions.}
For arbitrary score distributions, strong bounds for our problems seem elusive. But empirically significant cases often exhibit some type of smoothness (moderate change), except for point spikes -- score values repeated for a large fraction of positive or negative examples: ($i$) for classifiers with a limited range of possible output scores, ($ii$) when some inputs repeat verbatim many times, ($iii$) when a dominant feature value determines the score.
We call such score distributions {\em well-behaved}. 
Formally, 
we define a spike as any point with \red{probability mass}\eat{of positive or negative examples is} $>\phi$, hence at most $1/\phi$ spikes exist.\eat{in the score distribution function.}

\begin{definition}
\label{def:wellbehaved}
Let ${p}(s)$ and ${n}(s)$ denote the score (mass) functions $p$ and $n$ of the positive and negative examples respectively. 
Spikes are those points $s$ for which $p(s) > \phi$ or $n(s) > \phi$. 
We say that the score distribution is \textit{$(\phi, \ell)$-well-behaved} if it is $\ell$- Lipschitz between spikes. 
\begin{equation}
    |{p}(s) - {p}(s + \Delta) | \le \ell \Delta \quad \text{and} \quad 
    |n(s) - n(s + \Delta) | \le \ell \Delta
    \label{eq:lipschitz}
\end{equation}
\noindent
This smoothness condition for a parameter $\ell$ captures the idea that the amount of positive and negative examples does not change too quickly with $s$ (barring spikes in $[s, s + \Delta]$).
\end{definition}
\para{Balanced input assumption.}
\blue{For brevity, we assume that counts of positive and negative examples, $P$ and $N$ respectively, are bounded fractions (not too skewed) of the total number of examples $M=P+N$. 
Our results hold regardless of the balance, but the simplified proofs} expose core dependencies between results.
In some places, \blue{$P=N$ (perfect balance) maximizes our error bounds and illustrates} worst-case behavior. 
\eat{Although it is a reasonable assumption to make for many scenarios,} {\em Class imbalance} occurs in practice, \blue{e.g.,
 in recognizing hate speech, where $P\ll N$.}
\blue{We assume that classifier accuracy is\eat{lower-bounded by a constant} $>0.5$ (otherwise negating classifier output improves accuracy). Hence, for balanced inputs,} the fraction of true positives is at least a constant.



\para{Calibration.}
Given a score function $s$, calibration defines a transformation to apply to $s$ to obtain an accurate estimate of the probability that the example is positive.
That is, for a set of examples and labels $(x_i, y_i)$, we want a function $c()$, so that $c(s(x_i)) \sim \Pr[ y_i = 1]$. 
There are many approaches to find such a mapping $c$, such as isotonic regression, or fitting a sigmoid function. 
A baseline approach \red{is to perform} histogram binning on the function, with buckets chosen based on quantile boundaries. 
More advanced approaches combine information from multiple histograms in parallel~\citep{NaeiniCH15}. 
To measure the calibration quality, expected calibration error (ECE) arranges the predictions for a set of test examples into a fixed number of bins and computes the expected deviation between the true fraction of positives and predicted probabilities for each bin\footnote{\red{Formally, the ECE is defined  by \citet{NaeiniCH15} as 
$\sum_{j=1}^{K} P(j) |o(j) - e(j)| $, where $P(j)$ is the (empirical) probability that an example falls in the $j$th bucket (out of $K$), 
while $o(j)$ is the true fraction of positive examples that fall in the $j$th bin, and $e(j)$ is the fraction predicted by the model.}}. 

\para{Precision, Recall, Accuracy.}
Given a classifier that makes (binary) predictions $\pred(x_i)$ of the ground truth label $y_i$ on $M$ examples $x_i$, standard classifier quality metrics include:
\begin{itemize}
\vspace*{-2mm}
\itemsep0mm
    \item 
    \textbf{Accuracy}: the fraction of correctly predicted examples, i.e., $|\{ i : \pred(x_i) = y_i\}|/M$.
    \item
    \textbf{Recall}: the fraction of correctly predicted positive examples (a.k.a. true positive ratio), i.e., \newline
    \mbox{$|\{ i : \pred(x_i) = y_i = 1\}|/|\{ i : y_i = 1\}|$}.
    \item
    \textbf{Precision}: the fraction of examples labeled positive that are labeled correctly, i.e.,\newline
    \mbox{$|\{ i : \pred(x_i) = y_i = 1\}|/|\{ i : \pred(x_i) = 1\}|$}.
\end{itemize}
\para{ROC AUC} (Receiver Operating Characteristic Area Under the Curve) is often used to capture the quality of a trained ML classifier. 
Plotting TPR against FPR generates the ROC curve, and the ROC AUC (AUC for short) represents the area under this curve. 
The AUC can be computed in several equivalent ways.
Given a set of labeled examples $E = \{ (x, y) \}$ with $\pm1$ label $y$, the AUC equals the probability that a uniformly-selected positive example ($y=1$) is ranked above a uniformly-selected negative example ($y=-1$). 
Let $N$ be the number of negative examples, $|\{(x, -1) \in E\}|$, and $P=|\{(x, 1) \in E\}|$. 
Then \citep{Hand:Till:01}
\begin{equation} \textstyle \AUC = \frac{1}{PN} \sum_{(x, 1) \in E} \sum_{(z, -1) \in E} \mathbb{I}[s(x) > s(z)] 
\label{eq:auc}
\end{equation}
\blue{
This expression of AUC simplifies computation, but compares pairs of examples across clients.  
We avoid such distributed interactions by evaluating metrics via histograms.} 

\subsection{Privacy models}
\label{sec:privacymodels}
\para{Federated Privacy via Secure Aggregation (\fed)} assumes that
each client holds \red{one} (scalar or vector) value $x^{(i)}$.
\red{In practice, clients may hold multiple values, but this can easily be handled in our protocols by working with their sum.  
In what follows, we assume a single value since this represents the hardest case for ensuring privacy and security.
The Secure Aggregation protocol computes the sum,} $X = \sum_{i=1}^{\clients} x^{(i)}$ without revealing any intermediate values. 
Various cryptographic protocols provide distributed implementations of Secure Aggregation~\citep{secagg, BellBGK20}, or aggregation can be performed by a trusted aggregator, e.g., a server with secure hardware~\citep{Zhao:Jiang:Feng:Wang:Shen:Li:21}. 
While secure aggregation provides a practical and simple-to-understand privacy primitive, it does not fully protect against a knowledgeable adversary. 
In particular, knowing the inputs of all clients except for one, the adversary can\eat{infer the unknown value from the aggregated value, by} subtract the values that they already hold. 
Hence, DP guarantees \red{are sought for} additional protection.

\para{\red{Distributed Differential Privacy (\cdp)}.}
The model of differential privacy (DP) requires the output of a computation to be a sample from a random variable, so for two inputs that are close, their outputs have similar probabilities -- \red{in our case, captured as inputs that vary by the addition or removal of one item (event-level privacy)}~\citep{Dwork:Roth:14}. 
DP is often achieved by adding calibrated noise from a suitable statistical distribution to the exact answer: 
in our setting, this is instatiated by \red{introducing distributed noise at each client which sums to give discrete}
Laplace noise with variance $O(1/\epsilon^2)$.
The resulting (expected) absolute error for count queries is $O(1/\epsilon)$, and for means of \clients\ values is $O(1/\epsilon \clients)$. 
\red{The distributed noise is achieved by sampling from the difference of two (discrete) P\'{o}lya distributions~\citep{BalleBGN20}, which neatly
combines with secure aggregation so only the noisy sum is revealed}\footnote{\red{Other privacy noise is possible via Binomial~\citep{DworkKMMN06} or Skellam~\citep{Skellam} noise addition.}}. 

\para{Local Differential Privacy (\ldp).}
Local Differential Privacy can be understood as applying the DP definition at the level of an individual client: each client creates a message based on their input, so that whatever inputs the client holds, the probability of producing each possible message is similar.
For a set of $\clients$ clients who each hold a binary $b_i$, we can estimate the sum and the mean of $b_i$ under $\epsilon$-LDP by applying randomized response~\citep{RandResponse1965}. 
For the sum over $\clients$ clients, the variance is $O(\clients/\epsilon^2)$, and so the absolute error is proportional to 
$\sqrt{\clients}/\epsilon$.  
After rescaling by $\clients$, the variance of the mean is $O(1/(\clients\epsilon^2))$, and 
so the absolute error is proportional to $1/\epsilon\sqrt{\clients}$~\citep{survey1}. 
These bounds hold for histograms via ``frequency oracles'', when each client holds one out of $B$ possibilities --- we use Optimal Unary Encoding~\citep{wangetal} and
 build a (private) view of the frequency histogram, which can be normalized to get an empirical PMF. 

In what follows, we treat $\epsilon$ as a fixed parameter close to 1 (for both \ldp and \cdp cases). 

\subsection{Building score histograms}
\label{sec:scorehist}
A key step in our algorithms is building an equi-depth histogram of the clients' data. 
That is, given \clients\ samples as scalar values, we seek a set of boundaries that partition the range into $B$ buckets
with (approximately) equal number of samples per bucket.
Fortunately, this is a well-studied problem, so in Appendix~\ref{app:scorehist} we review how such histograms can be computed under each privacy model and analyze their accuracy as a function of parameters $h$ and $\epsilon$. 
\red{The novelty in our work is the way that we can use these histograms to provide classifier quality measures with privacy and accuracy guarantees across a range of federated models.}

For federated computation, we want to represent the distribution of scores with histograms. 
Based on a set of $B$ buckets that partition $[0,1]$, we obtain two histograms, $n$ and $p$, which provide the number of examples whose score falls in each bucket.  
Specifically, $p_i$ and $n_i$ give the number of positive and negative examples respectively whose score falls in bucket $i$. 
In what follows, we set the histogram bucket boundaries based on the (approximate) quantiles of the score function for the given set of examples $E$, so that 
$p_i + n_i \le (P + N)/B$, where $P$ and $N$ denote the total number of positive and negative examples respectively. 

\red{
\para{Overcoming heterogeneity.}
A common concern when working in the federated model is data heterogeneity: the data held by clients may be non-iid (i.e., some clients are more likely to have examples of a single class), and some clients may hold many more examples than others. 
By working with histogram representations we are able to overcome these concerns: the histograms we build are insensitive to how the data is distributed to clients, and the noise added for privacy is similarly indifferent to data heterogeneity.  
Thus we can state our results solely in terms of a few basic parameters (number of clients, number of histogram buckets etc.), independent of any other parameters of the data. 
}

\begin{table*}[t]
    \centering
    \caption{\label{tab:results}
    Simplified variants of error bounds derived in this paper for three privacy models.}
    \begin{tabular}{l|c|c|c}
         &  \fed & \cdp & \ldp \\
         \hline
         P/R/A (for a given classifier) & 0 
         & $1/\epsilon \clients$ 
         & $1/\epsilon{\clients}^{1/2}$ 
         \\
         P/R/A (for a score function and varying threshold) & $1/B$ 
         & $1/\epsilon^{2/3}\clients^{2/3}$ 
         & $1/\epsilon^{2/3} \clients^{1/3}$ 
         \\
         ROC AUC & $1/B^2$ 
         & $(\frac{1}{\epsilon} + \frac{1}{B})\frac{1}{\clients}$  
         & $h/\epsilon{\clients}^{1/2}$ 
         \\
         Expected Calibration Error (ECE) & $1/\clients^{1/3}$ 
         & $1/\clients^{1/3}$
         & $1/\epsilon^{1/2} \clients^{1/4}$ 
    \end{tabular}
\end{table*}

\subsection{Summary of our results}
\label{sec:summary}
Table~\ref{tab:results} presents simplified versions of our main results.
Here and throughout, we express error bounds in terms of the expected absolute error, which can also be used as a bound that holds with high probability via standard concentration inequalities.  
\blue{Without requiring any i.i.d. assumptions for distributed clients}, our results clarify the expected magnitude of the error, which should be small in comparison to the quantity being estimated: tightly bounding the error values ensures accurate results. 
All the estimated quantities are in the $[0, 1]$ range, and for most (binary) classifiers of interest the four quality metrics \red{will be} $\geq \frac12$. 
For calibration, the expected calibration error is a small value in [0, 1].

To keep the presentation of these bounds simple,  we make the {\em balanced input assumption} (from Section~\ref{sec:defs}), i.e., there are $\Theta(\clients)$ positive examples and $\Theta(\clients)$ negative examples among the $\clients$ clients. 
Error bounds are presented as a function of the number of clients, $\clients$, the privacy parameter $\epsilon$, 
the number of buckets used to build a score histogram, $B$, and the height of the hierarchy used, $h$ \red{(see Section~\ref{sec:pra-score})}. 
Across the various problems the error increases as we move from 
\fed to \cdp to \ldp. 
This is expected, as the noise added in each case increases to compensate for the weaker trust model. 

Other trends we see are not as easy to guess. 
Increasing the number of buckets $B$ often helps reduce the error, but this is not always a factor, particularly for the \ldp results. 
Increasing the number of examples, $\clients$, typically decreases the error, although the rate of improvement as a function of $\clients$ varies from $1/\clients^{1/4}$ in the worst case to $1/\clients$ in the best case. 
Our experimental \red{findings}, presented in Appendix~\ref{app:expts}, agree with this analysis and
confirm the anticipated impact of increasing $\clients$ and of varying the parameters $B$ and $h$. 
We observe high accuracy in the \fed case and good accuracy when DP noise is added.  
Calibration error for \cdp is insensitive to $\epsilon$ as explained after Theorem \ref{thm:calib:dp}.
These results help building full-stack support for practical federated learning, and show the practicality of federated classifier evaluation.

\section{Federated Computation of Precision, Recall, Accuracy}
\label{sec:pramain}
In this section, we discuss how to approximate the precision, recall and accuracy of a classifier in a federated setting. 
As a warm-up, we consider the case when the classifier is fully specified; this can be handled straightforwardly by federated computation of basic statistics. 
However, when the threshold $T$ is specified at query time, we need to make use of score histograms to gather the necessary information to provide accurate answers. 

In what follows, we bound the accuracy of the estimates
(all detailed proofs are given in Appendix~\ref{app:proofs}).

\begin{theoremEnd}{fact}
\label{fact:ratio}
Assume $\hat{A} = A \pm \alpha$ and $\hat{G} = G \pm \gamma$, 
where $\gamma \leq G/2$ and $\hat{X} = X\pm x$ is shorthand for $\hat{X} \in [X-x, X+x]$.
When estimating a fraction $\frac{A}{G} \leq 1$,
we have
$|\frac{\hat{A}}{\hat{G}} - \frac{A}{G}| =O(\frac{\alpha+\gamma}{G})$. 
\end{theoremEnd}

\begin{proofEnd}
\begin{align}
    \left| \frac{A \pm \alpha}{G \pm \gamma} - \frac{A}{G} \right|
    & =
    \left| \frac{A \pm \alpha}{G(1 \pm \gamma/G)} - \frac{A}{G}\right| \nonumber
    \\& = 
    \left| \frac{(A \pm \alpha)(1 \mp 2\gamma/G)}{G} - \frac{A}{G}\right| \nonumber
    \\& \le 
    \frac{2A\gamma}{G^2} + \frac{\alpha}{G} + \frac{2\alpha\gamma}{G^2}
    \\& = O\left(\frac{\alpha + \gamma}{G}\right)
    \label{eq:approxbound}
\end{align}
where we use $A/G \leq 1$ and $\gamma/G \leq 1/2$ to simplify in the final step. 
\end{proofEnd}

\subsection{\red{Warm-up:} Precision, recall and accuracy for a given classifier}

In the federated setting, each client holds labeled examples. When computing precision, recall, and accuracy ,it helps that the client knows whether a given example contributes to the numerator and/or the denominator of each metric. 
Such counts can be \red{collected} exactly in the secure aggregation case.
For \red{distributed and} local differential privacy noise, the estimation error is slightly more complicated, but still bounded. 
In all cases, we \red{can easily} bound the expected absolute error of the estimate, \red{and show these bounds for comparison with subsequent cases}. 

\begin{theoremEnd}{lemma}
Under the balanced input assumption,
for a given classifier, one can build
federated estimators of precision, recall and accuracy
as follows:
  \begin{itemize}
      \item 
      In the basic \fed case we can compute the exact result without noise;
      \item
      In the \cdp case,
      the error is $O(1/\epsilon \clients)$; 
      \item 
      In the \ldp case,
      the error is $O(1/\epsilon \sqrt{\clients})$.
  \end{itemize}
\end{theoremEnd}

\begin{proofEnd}
We consider the cases of precision, recall and accuracy in turn:

\textbf{Accuracy.} 
Accuracy is most straightforward, as we can assume that we know the total number of examples $M$ exactly.  
Then we can collect binary responses from client $i$ on whether or not $\pred(x_i) = y_i$.  
In the \fed case, as with the others, we can gather these counts exactly from the clients. 
Under $\epsilon$-\ldp noise, the additive error on the estimated accuracy behaves as 
$\frac{1}{\epsilon\sqrt{\clients}}$, while for \cdp it is
$\frac{1}{\epsilon \clients}$, from bounds in Section~\ref{sec:privacymodels}. 

\textbf{Recall.}
Recall requires us to know $P$, the total number of positive examples, and $TP$, the number of true positives from the training data. 
Suppose we can estimate both $P$ and $TP$ with additive error at most $\alpha = \gamma < P/2$. 
Then we can use Fact~\ref{fact:ratio} to bound the error as
$O(\gamma/P)$. 
We now argue that this error can be obtained in our two DP models. 

In the \cdp case, \red{the aggregated discrete Laplace noise} addition ensures that 
$\alpha = \gamma = O(1/\epsilon)$ (with high probability).
For \ldp, we have that $\gamma = \sqrt{\clients}/\epsilon$, which, provided that $\sqrt{\clients}/\epsilon < P/2$, gives the expected absolute error as 
$O(\frac{\sqrt{\clients}}{\epsilon P})$. 
In the case when the number of positive examples $P$ is a constant fraction of the total number of examples $\clients$, this error bound simplifies to $O(1/\epsilon\sqrt{\clients})$. 

\textbf{Precision.}
The analysis for precision is similar.  
We estimate $TP$ and $(TP + FP)$ with additive error $\alpha = \gamma$ to obtain 
additive error $O(\gamma/(TP + FP))$ via Fact~\ref{fact:ratio} again, 
when $\gamma < (TP + FP)/2$. 
Under the balanced input assumption, 
the classifier must classify a constant fraction of examples as positives provided the classifier accuracy is at least $\frac12$, say.
Then we can simplify this bound to 
$O(1/\epsilon\sqrt{\clients})$ in the \ldp case, and 
$O(1/\epsilon \clients)$ for \cdp. 
\end{proofEnd}

\subsection{Precision, recall and accuracy for a score function}
\label{sec:pra-score}

Next, instead of a specified binary classifier, we only have a score function $s(\cdot)$ with values in [0, 1] for each example $x$. We seek statistics that would help estimate the precision, recall and accuracy of the classifier defined by a threshold $T$, where $T$ can be chosen at query time. Our solution makes use of score histograms (Section~\ref{sec:scorehist}) built over the positive and negative examples. \red{The idea is to break the calculation into a discrete sum over histogram buckets, where we can bound the uncertainty due to this discretization by limiting the number of examples in the bucket, and bound the uncertainty due to privacy noise by limiting the number of buckets. 
The final results come by balancing these two sources of uncertainty to determine the optimal number of buckets as a function of privacy $\epsilon$ and number of client values $\clients$.}

\begin{theoremEnd}{thm}
\label{thm:pra}
Given a score histogram for positive and negative examples built based on a hierarchy of height $h$, 
we can compute estimates for precision, recall and accuracy based on a threshold $T$ which approximate
the true precision, recall and accuracy for a threshold $T' \in T \pm \Delta$, under the balanced input assumption, as follows: 
\begin{itemize}
    \item 
    In the basic \fed case, we achieve error $O(1/B)$ with $\Delta = 2^{-h}$;
    \item
    For \cdp, we achieve error $O(1/(\epsilon\clients)^{2/3})$ with $\Delta = O(h^{3/2}/\epsilon \clients + 2^{-h})$;
    \item
    For \ldp, we achieve error $O(1/(\epsilon^{2/3}\clients^{1/3}))$ with $\Delta = O(h/\epsilon\sqrt{\clients} + 2^{-h})$.
\end{itemize}
\end{theoremEnd}

\begin{proofEnd}
All our approaches are based on using the score histogram of positive and negative examples. 
Per Appendix~\ref{app:scorehist}, we can compute such a histogram that provides an answer that is accurate up to a small uncertainty in $T$, which we write as $(T \pm \Delta)$. 
The histogram is based on a parameter $h$ that determines the height of the hierarchy used to construct it. 
Under the \fed model, we have that $\Delta =  2^{-h}$, whereas for
\cdp $\Delta = O(h^{3/2}/\epsilon \clients + 2^{-h})$ and for 
\ldp $\Delta = O(h/\epsilon \sqrt{\clients} + 2^{-h})$, as explained in Section~\ref{sec:scorehist}. 
We now consider each classifier metric in turn.

\para{Accuracy.}
Accuracy is the easiest function to handle, since we just have to compute (for the numerator)
\begin{equation}
    | \{ i : y_i = -1 \wedge \pred(x_i) < T\}| + |\{ i : y_i = 1 \wedge \pred(x_i) \ge T \}|
\end{equation}

That is, the number of negative examples with a score below $T$ plus the number of positive examples with a score of at least $T$.
We can estimate both these quantities with additive error at most $1/B$ using a $B$-bucket equi-depth histogram (without noise addition).

In the \cdp case, 
\red{there is discrete Laplace noise on each} bucket count to mask the presence of any individual. 
We can bound the error from this noise to be of order
$\frac{\sqrt{B}}{\epsilon \clients}$, by summing variances, giving a total error bound of
$O(\frac{1}{B} + \frac{\sqrt{B}}{\epsilon \clients})$. 
We can balance these two terms so that
$\frac{1}{B} = \frac{\sqrt{B}}{\epsilon \clients}$, 
so $B^3 = \epsilon^2\clients^2$, i.e.,
$B = (\epsilon \clients)^{2/3}$. 
This gives the total error as 
$O(1/(\epsilon \clients)^{2/3})$. 

Under $\epsilon$-\ldp noise, we obtain an additional error term of $\frac{\sqrt{B}}{\epsilon \sqrt{\clients}}$
(by summing the variance over $B$ buckets).
Balancing these two terms means we should set
$1/B = \sqrt{B}/{\epsilon\sqrt{\clients}}$, 
i.e., 
$B^3 = \epsilon^2 \clients$, and so 
$B = \epsilon^{2/3} \clients^{1/3}$. 
Under this setting, the total error is bounded as
$O(1/\epsilon^{2/3}\clients^{1/3})$. 

\para{Recall.}
The same histogram approach works for recall. 
Using a histogram, we aim to estimate the number of true positives, which is the number of positive examples
above the threshold, divided by the total number of positives. 
Without DP noise, we can compute $P$, the number of positive examples, exactly for the denominator, but we 
incur error $M/B$ for the numerator, giving error $M/BP$. 
To simplify this expression, we can invoke the {\em balanced input} assumption, which bounds this by $O(1/B)$ since $M = O(P)$. 

Including \ldp noise, we incur error $\alpha = \sqrt{B\clients}/\epsilon$ when summing over $B$ buckets.
We also have error $\gamma = \sqrt{\clients}{\epsilon}$ for estimating $P$. 
Using Fact~\ref{fact:ratio}, the error is dominated by 
$O(\sqrt{B\clients}/\epsilon P + \clients/BP)$.  
This is again balanced by setting $B = \epsilon^{2/3}\clients^{1/3}$.
Likewise, for \cdp noise, we have $\alpha = \sqrt{B}\clients/\epsilon$ and $\gamma = \clients\epsilon$, 
which leads to choosing $B = (\epsilon\clients)^{2/3}$ for total error
$O(1/B) = O((\epsilon\clients)^{-2/3})$ under the balanced input assumption. 

\para{Precision.}
The bounds for precision are similar. 
Using a histogram, we want to first count how many examples are correctly classified as positive -- this is the number of positive examples above the threshold $T$. 
We scale this by the total number of examples that are classified as positive, which is just the number of examples above the threshold $T$. 
Under secure aggregation, we can estimate both of these with error $M/B$, which is due to the histogram bucketing. 
Plugging these into \eqref{eq:approxbound}, the error bound is 
$O(M/B(TP+FP))$.
Under our {\em balanced input assumption} that a constant fraction of the examples are positives, and that the classifier has at least a constant accuracy, we can simplify this bound to $O(1/B)$.

With \ldp noise, we incur additional noise of $\alpha = \gamma = \frac{\sqrt{BM}}{\epsilon}$
on both these these quantities.  
Now the error bound is $O(\frac{1}{(TP + FP)} (\frac{M}{B} + \frac{\sqrt{BM}}{\epsilon\sqrt{M}}))$. 
Once more, balancing this error leads us to pick 
$B = \epsilon^{2/3}M^{1/3}$, 
which gives an error of the form $O(M^{2/3}/\epsilon^{2/3}(TP+FP))$. 
If $FP + TP$ is a constant fraction of $M$, then we simplify this to 
$O(1/\epsilon^{2/3} M^{1/3})$. 

Similarly for \cdp noise, 
we have
$\alpha = \gamma = \frac{\sqrt{B}M}{\epsilon}$, which leads to 
error $O(1/(\epsilon M)^{2/3})$ under the same assumptions on positive examples. 

Combining these bounds with the error introduced by using a score histogram to find the bucket boundaries on the threshold as $\Delta$, we obtain the results stated in the theorem. 
\end{proofEnd}

We observe that in the \fed case, accuracy improves without limit if we increase the height of the hierarchy $h$ arbitrarily and scale $B\sim 2^h$.
The only cost is that the resulting histogram built by the aggregator is $O(2^h)$ in size. 
However, for \cdp and \ldp, increasing $h$ increases the imprecision $\Delta$: there is uncertainty due to the privacy noise, which eventually outweighs the fidelity improvement due to smaller histogram buckets. 
Our analysis in the proof of Theorem~\ref{thm:pra} balances the two terms to find a near-optimal setting of $B$ that yields the stated bounds.  

\section{Federated Computation of ROC AUC}
\label{sec:rocauc-bounds}

Estimating ROC AUC is a fundamental problem in classifier evaluation. 
It is particularly challenging in the federated setting, since it requires comparing how different examples are handled by the classifier, whereas these examples are usually held by different clients.  
However, it turns out that we can get accurate approximations of AUC without requiring communication amongst clients. 

We will use $B$-bucket score histograms (Section~\ref{sec:scorehist}) to define a histogram-based estimator for AUC:
\begin{equation}
\textstyle
    H_B = \frac{1}{PN} \sum_{i \in [B]} \left( \sum_{j < i} p_i n_j + \frac12 p_i n_i \right)
\label{eq:histauc}
\end{equation}
Recall that $P$ and $N$ denote the total number of positive and negative examples, 
while $p_i$ and $n_i$ denote the number of positive and negative examples in bucket $i$ of the histogram. 

We start by analysing the accuracy when the histogram contains exact counts, i.e., in the \fed case. 
Compared to the precise AUC computation, our uncertainty in this estimate derives from the $p_i n_i$ term: 
for any $j<i$, we know that all the pairs of items that contribute to $p_i n_j$ would be counted by~\eqref{eq:auc}, 
while for $j>i$, no pairs corresponding to $p_i n_j$ should be counted. 
However, within bucket $i$, we are uncertain whether all positive items are ranked higher than all negative items 
(in which case we should count $p_i n_i$ towards \eqref{eq:auc}), or 
vice-versa (yielding a zero contribution). 
The choice of $\frac12 p_i n_i$ in \eqref{eq:histauc} takes the midpoint between these two extremes. 
Later, we show that this is a principled choice for well-behaved score distributions. 

\subsection{Worst-case bounds via score histograms}
We first present a general bound on AUC estimation using a score histogram. 
A key insight is that, in the \fed case, the only uncertainty comes from the contribution to the AUC of positive and negative examples that fall in the same bucket.
Using an equi-depth histogram bounds the number of such examples, and so the absolute error drops as the number of buckets in the histogram grows.

\label{sec:aucbasic}
\begin{theoremEnd}{lemma}
\label{lem:auc}
In the \fed case, the additive error in AUC estimation  with a $B$-bucket score histogram is $O(1/B)$. 
\end{theoremEnd}
\begin{proofEnd}
In order to bound the error in our estimate of AUC, we can choose the histogram buckets based on the quantiles of the score function (Section~\ref{sec:scorehist}), so that $p_i + n_i = (P+N)/B$.  
The error in our estimate is at most 
$\frac{1}{2PN} \sum_{i \in [B]} p_i n_i$. 
We observe that this error is maximized when $p_i = n_i = (P + N)/2B$, so that the resulting (absolute) error
is 
\begin{align*}
\frac{1}{2PN} \sum_{i \in [B]} p_i n_i 
  & = 
    \frac{1}{2PN} \sum_{i \in [B]} \frac{(P + N)^2}{(2B)^2} 
 \\& = \frac{4}{2(P+N)^2} \frac{B (P+N)^2}{4B^2} 
 \\& = \frac{1}{2B}
\end{align*}
\noindent
using that our worst-case setting of $p_i$ and $n_i$ entailed that $P = N = (P+N)/2$ (the perfectly balanced input case). 
\end{proofEnd}

The proof considers worst-case allocations of examples with a uniform share of positive and negative examples in each bucket, yielding error exactly $1/2B$.
That is, using a histogram with $B$ buckets set by the quantiles of the score function suffices to bound the (additive) AUC error by $1/2B$. 
For instance, $B=50$ buckets ensure that the error  $\leq 0.01$. 
For a classifier with $\AUC > \frac12$, this bound promises a relative error of at most $1/B$. 

Despite similarities to the bounds for precision, recall and accuracy (Section~\ref{sec:pra-score}), 
this analysis for AUC is quite pessimistic. 
First, it assumes that all buckets have an equal number of positive and negative examples. 
For a realistic classifier, we would expect 
mostly negative examples in buckets with low scores, and mostly positive examples in
buckets with high scores. However, this property alone is insufficient to significantly improve the bound. 
For instance, if we assume that bucket $i$ has an $i/B$ fraction of positive examples and a $(1-i/B)$ fraction of negative examples, then calculation leads to bound of 
$1/3B$ on the error (not much improved).
In our experimental study (Appendix~\ref{app:expts}), the observed values of $\sum_{i \in [B]} (p_i n_i)$ are close to $1/3B$, implying that any analysis based solely on this quantity cannot improve the form of this bound. 

\subsection{Improved AUC error bound for well-behaved distributions}
\label{sec:aucbetter}
The worst-case bound allows extreme cases where all positive examples in a bucket are ranked above all negative examples in the same bucket, or vice-versa. 
We derive a tighter bound when the distribution functions of positive and negative examples 
are {\em well-behaved} as per Section~\ref{sec:defs}. 

\begin{theoremEnd}{thm}
\label{thm:aucbetter}
For inputs meeting the $(1/B, \Theta(1))$-well-behaved definition, the \fed error for AUC estimation using score histograms is $O(1/B^2)$. 
\end{theoremEnd}

\begin{proofEnd}
We consider the maximum uncertainty we can have within a single histogram bucket $i$ when the score distribution is $(1/B, \Theta(1))$-well-behaved. 
First, suppose that there is a spike within the bucket.  
Choosing our spike parameter $\phi = 1/B$, we have that the bucket must contain \emph{only} this spike, otherwise the bucketing would violate the quantile property.  
Thus, we have no uncertainty as to the contribution of the single point $x$ in this bucket to the AUC, as it is zero according to \eqref{eq:auc}. 
Hence, providing the $(\phi,\ell)$-well-behaved property holds for $\phi = 1/B$, we incur no error due to spikes. 

This leaves only buckets without spikes, which are then assumed to obey the Lipschitz condition with parameter $\ell = \Theta(1)$. 
Abusing notation slightly, let $p_i$ and $n_i$ denote the \red{mass} of positive and negative examples at the left hand end of the bucket. 
We reparameterize the \red{mass} function within a bucket based on a parameter $0 \leq \alpha \leq 1$, so that 
$p(\alpha)$ and $n(\alpha)$ give the \red{mass} of examples within the bucket at the point that is an $\alpha$ fraction across the bucket (from left to right). 
We define
  $L = \ell \Delta_i$, where $\Delta_i$ is the width of bucket $i$. 
Based on the above definitions, we have 
\[
| p(\alpha) - p_i | \leq L \alpha  \qquad \text{and} \qquad |n(\alpha) - n_i | \leq L \alpha
\]

Rearranging, we can write $p(\alpha) \in p_i \pm L\alpha$ and
$n(\alpha) \in n_i \pm L\alpha$. 

The contribution to AUC from this bucket is then bounded by integration of these linear bounding functions: 
\begin{align*}
\int_0^1 p(\alpha)  \int_0^\alpha n(\alpha') d\alpha' d\alpha
 & \in 
\int_0^1 (p_i \pm L\alpha) \int_0^\alpha(n_i \pm L\alpha') d\alpha' d\alpha  \\
& \in \int_0^1 (p_i \pm L\alpha) \left(n_i \alpha \pm \frac{L\alpha^2}{2} \right)d\alpha \\
& \in \left[ \frac{p_i n_i\alpha^2}{2} \pm \frac{L n_i \alpha^3}{3} \pm \frac{p_i L \alpha^3}{6} \pm \frac{L^2 \alpha^4}{8}   \right]^1_0 \\
& \in \frac12{p_i n_i} \pm \frac13 Ln_i \pm \frac16 Lp_i \pm \frac18 L^2 . 
\end{align*}
 
If we use $p_i n_i/2$ as our estimate of the contribution to the AUC from bucket $i$, the absolute error in this estimate is at most $Ln_i/3 + Lp_i/6 + L^2/8 = O(\Delta_i(p_i + n_i) + \Delta_i^2)$ (treating the Lipschitz parameter $\ell$ as a constant).
Without loss of generality, we can assume that $\Delta_i \leq 1/B$ -- the width of any bucket is at most $1/B$.  
Although this is not directly implied if we define the buckets by the quantiles of the score functions, we can additionally enforce this property without changing that there are $O(B)$ buckets in the histogram.  
Then the uncertainty in AUC contribution is $O((p_i + n_i)/B + 1/B^2) = O((P + N)/B^2)$, from our definition of bucket boundaries and the bound on the number of points in a bucket. 
Summed over all buckets, and normalized by the factor of $1/PN$, the absolute error in AUC is bounded by 
$O(\frac{P + N}{BPN}) = O(1/B\clients)$, 
under the {\em balanced input assumption}. 

To express this solely in terms of $B$, we can observe that we must have 
$B < N$ and $B < P$ (otherwise, we have empty buckets, which do not contribute to the error), and so 
the error bound is $O(B^{-2})$. 
\end{proofEnd}

This improved $1/B^2$ scaling is strong and produces tight error bounds for small $B$.
Picking $B\approx 100$ gives error $\approx 10^{-4}$, small enough for most conceivable applications. 
\eat{It remains to determine whether the well-behaved property holds in practice.}
Empirically we observe that absolute errors in AUC estimates closely follow $O(B^{-2})$ on test data (Appendix~\ref{app:expts}).

\subsection{AUC noise addition for \cdp} 
\label{sec:auccdp}
When we consider noise for differential privacy, we now have to weigh the cost of noise in every bucket. 
This quickly becomes the dominant cost. 

\begin{theoremEnd}{thm}
\label{auc:cdp}
\label{thm:auccdp}
Under \cdp, the AUC estimation error bound with the {\em balanced input assumption} is 
$O((\frac{1}{\epsilon} + \frac{1}{B}) \frac{1}{M})$.
\end{theoremEnd}

\begin{proofEnd}
Under differential privacy, we additionally have to account for privacy noise on the counts. 
We first consider the effect of centralized DP noise added to each histogram bucket. 
Recall that, as described in Section~\ref{sec:privacymodels} the effect of the $\epsilon$-DP noise is to add unbiased noise of 
variance $O(1/\epsilon^2)$ 
(i.e., with magnitude $\Theta(1/\epsilon)$) to every count. 
This means that there are errors introduced in the estimates of $p_i n_j$, as well as $p_i n_i$. 

Errors also arise due to the variation in the size of histogram buckets: 
if we estimate quantiles under differential privacy, then we no longer guarantee that there are exactly $\clients/B$ examples in each histogram bucket. 
However, the analysis is not highly sensitive to this issue, and it suffices to assume that the private histogram guarantees that there are between $\clients/2B$ and $2\clients/B$ examples in each bucket. 
This would be the case using the \cdp histogram construction from Section~\ref{app:scorehist} for typical choices of the parameters $h$, $\epsilon$, $\clients$ and $B$ (say, $\clients$ more than a few hundred). 

We can make use of the expression for the variance of the product of two independent random variables, 

\begin{equation}  \Var[XY] = \Var[X]\Var[Y] + \Var[X] (\E[Y])^2 + \Var[Y] (\E[X])^2 .
\label{eq:varproduct}
\end{equation}

We apply this expression to the estimate of $p_i n_j$, since the random variables representing privacy noise on each of $p_i$ and $n_j$ are truly independent. 
The total variance in the use of a noisy histogram with $B$ buckets, $\hat{H}_B$, in \eqref{eq:histauc} to approximate \eqref{eq:auc} is given by
\begin{align*}O\left(\sum_{i, j} \frac{1}{\epsilon^4} + \frac{p_i^2}{\epsilon^2} + \frac{n_j^2}{\epsilon^2}  \right)
& = O\left(  \frac{B^2}{\epsilon^4} + B\sum_{i \in [B]} \frac{p_i^2 + n_i^2}{\epsilon^2}\right) \\&
 = O\left( \frac{B^2}{\epsilon^4} + \frac{P^2 + N^2}{\epsilon^2}\right)
\end{align*}

This expression is dominated by
the quadratic term in $P$ and $N$ for $\epsilon$ at least a constant,
i.e., we can use 
$O( (P+N)^2 \left( \frac{1}{\epsilon^2}\right) )$ as a bound on the variance, since we can assume $P > B$ and $N>B$. 
Combining the error bound from Theorem~\ref{thm:aucbetter}
and after normalizing by the factor of $PN$, this 
yields an absolute error of magnitude
$O( (\frac{1}{\epsilon} + \frac{1}{{B}}) \frac{P+N}{{PN}})$, i.e., augmenting $1/{B}$ from the noiseless case with an additional 
 $1/\epsilon$. 
 Under the {\em balanced input assumption}, we can write the total error bound as 
 $O((\frac{1}{\epsilon} + \frac{1}{B}) \frac{1}{\clients})$.
 \end{proofEnd}
 
 That is, the error is comprised of two components: privacy noise of $O(1/\epsilon \clients)$, and ``bucketization'' noise of $O(1/B\clients)$. 
  Since $\epsilon$ can usually be treated as fixed, this rules out asymptotic benefit for increasing $B$ above $\Theta(\epsilon)$: when $B$ is large enough, the error due to privacy noise will dominate, and using more buckets will not help. 
  Empirical data in Appendix~\ref{app:aucexpts}
  confirms this: for $\epsilon=1$, $B\gg 20$ makes negligible difference in terms of accuracy. 

\subsection{AUC noise addition for \ldp}
The \ldp case is similar, except the magnitude of the noise is larger, since we are incurring noise on every example. 
Here, the error of the quantile estimates will also be larger, but this does not impact the calculation. 

\begin{theoremEnd}{thm}
\label{auc:ldp}
\label{lem:aucldp}
\label{thm:aucldp}
Under \ldp, the error bound for AUC estimation with the {\em balanced input assumption} using a hierarchy of height $h$ is $O(h/\epsilon\sqrt{\clients})$.
\end{theoremEnd}

\begin{proofEnd}
As in the \cdp case, we assume that $\clients$ is large enough that the error from determining the quantile boundaries is small enough that each bucket has a constant multiple of $\clients/B$ examples in it. 
This means that $h/\epsilon\sqrt{\clients} \leq \clients/B$. 
Rearranging, we require $Bh/\epsilon = O(\clients^{3/2})$. 
For constant $\epsilon$, and typical bounds $h \leq 20$, $B \leq 10^3$, we have
that this will hold provided that $\clients$ is in the order of thousands or more. 

The variance of a range query on a population of size $n$ will be $n V_\epsilon$, where $V_\epsilon$ denotes the variance from the LDP frequency oracle. 
For example, when we use a hierarchical histogram~\citep{Cormode:Kulkarni:Srivastava:19} of height $h$ with
Optimal Unary Encoding, $V_\epsilon = \frac{h^2 \exp(\epsilon)}{(\exp(\epsilon) - 1)^2}$. 

As usual, we write $\clients = N + P$ to denote the total number of examples.  
Considering the variance in the estimation of $\sum_{i \in [B]} \sum_{j < i} p_i n_j$ via~\eqref{eq:varproduct}, we obtain
\begin{align*}
    \Var[ \sum_{i \in [B]} \sum_{j < i} p_i n_j ]
&  =     
  O\left(\sum_{i,j} M^2V_\epsilon^2 + p_i^2 M V_\epsilon + n_j^2 M V_\epsilon \right) \\
& = 
O\left( \sum_{i,j} M^2 V_\epsilon^2 + M^3 V_\epsilon/B^2 \right) \\
& = 
O\left( M^2 B^2 V_\epsilon^2 + M^3 V_\epsilon \right)
\end{align*}

For small $B$, the term in $M^3$ will dominate. 
If we balance the two terms, we obtain 
$B = O(\sqrt{\clients/V_\epsilon})$. 

For $\epsilon = O(1)$, we have that $V_\epsilon = O(h^2/\epsilon^2)$.
Consequently, the absolute error is of magnitude
$O(\sqrt{V_\epsilon/\clients}) = O(\frac{h}{\epsilon\sqrt{\clients}})$.
That is, the dependence on $\clients = P+N = O(\sqrt{PN})$ is weakened, and so the error decreases more slowly as the number of examples is increased. 
\end{proofEnd}

We can compare this bound to a result of~\citet{BellBGK20} where 
a bound of $O(\frac{h^{3/2}}{\epsilon\sqrt{M}})$
is derived for LDP AUC estimation.
Their setting assumes a discrete domain with $2^h$ possible values and non-private classes of the examples, whereas we remove those assumptions. 

\begin{table}[t]
\centering
\caption{\label{tab:exp_setup}
Data and classifiers from \red{three different} ``tabular playground'' Kaggle competitions used for evaluation.
}
\begin{tabular}{c|c|c|c}
\sc Kaggle & \sc Data & \sc Baseline & ROC \\
\sc challenge & \sc rows & \sc classifier & AUC\\
\hline
    Sep 2021 & 958K & LightGBM & 0.79\\
    Oct 2021 &  1M & XGBoost & 0.85\\
    Nov 2021 & 600K & Logistic Regr. & 0.73\\ 
\end{tabular} 
\end{table}

\begin{figure*}[th]
\centering
\includegraphics[trim=800 25 800 25,clip,width=0.59\textwidth]{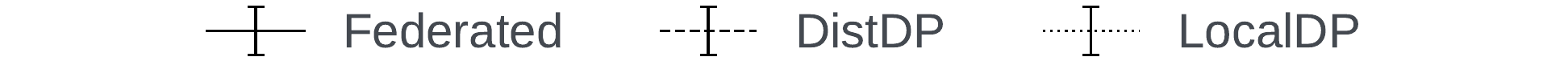}\newline
\hspace*{-9mm}
\subcaptionbox{
September 2021 data\label{auc-sep-pop}}
{\includegraphics[width = \figwidth]{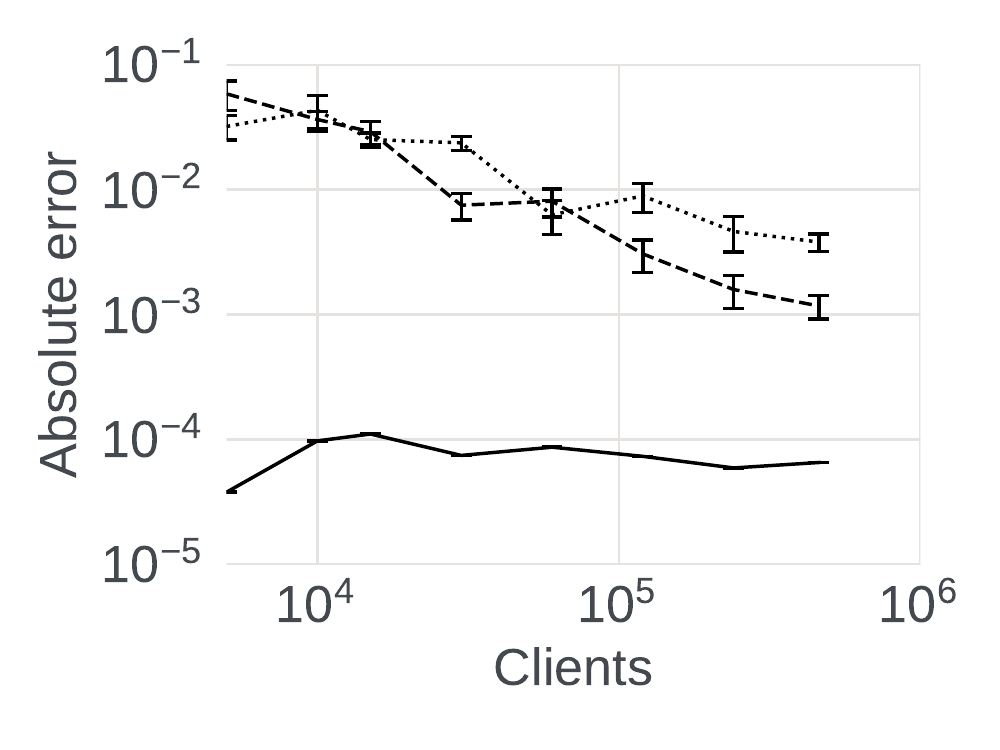}}%
\subcaptionbox{
October 2021 data\label{auc-oct-pop}}
{\includegraphics[width = \figwidth]{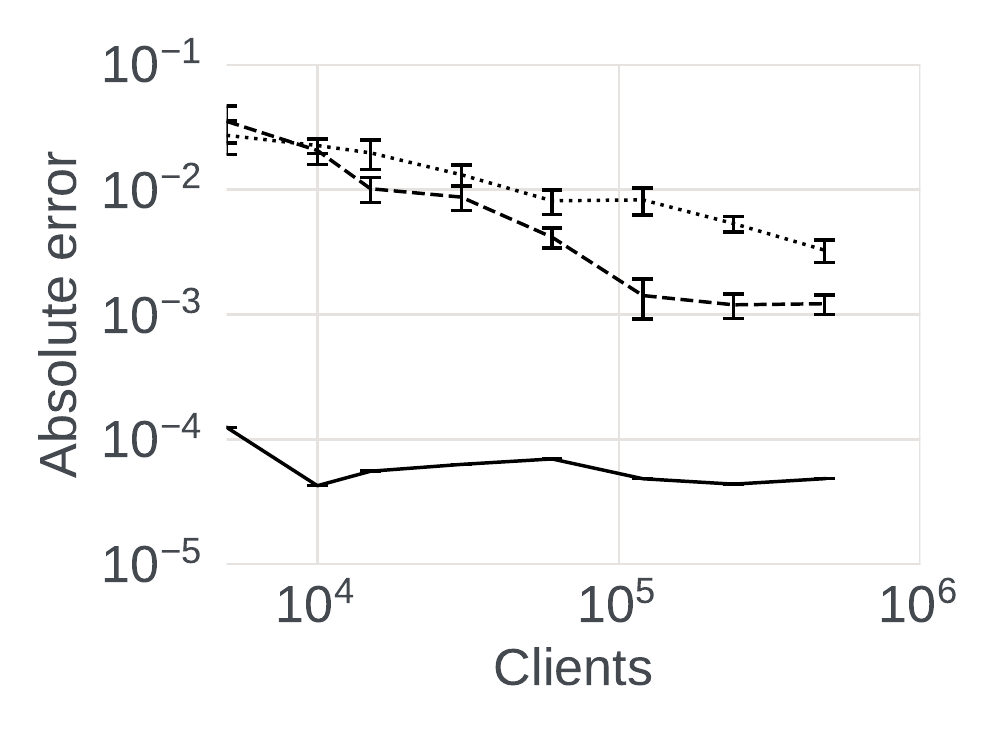}}%
\subcaptionbox{
November 2021 data\label{auc-nov-pop}}
{\includegraphics[width = \figwidth]{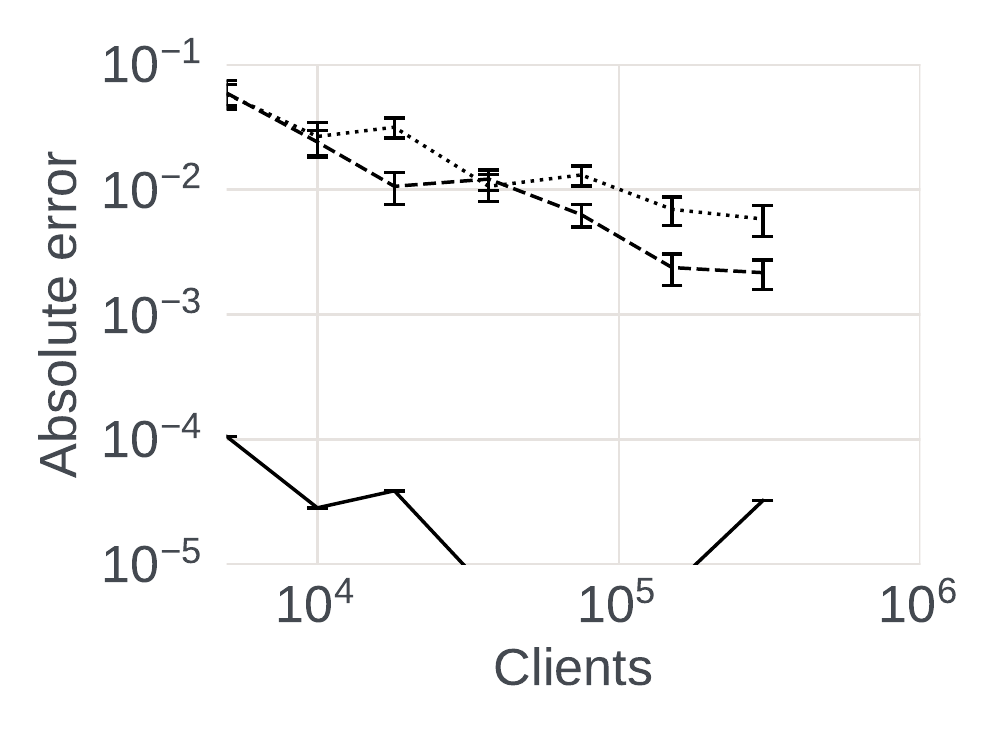}}
\caption{Accuracy for ROC AUC estimation with varying population size}
\label{fig:auc-pop}
\end{figure*}

\begin{figure*}[!t]
\centering
\includegraphics[trim=800 25 800 25,clip,width=0.59\textwidth]{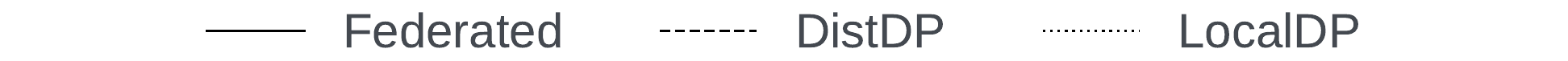}\newline
\hspace*{-9mm}
\subcaptionbox{
September 2021 data\label{calib-sep-pop}}{\includegraphics[width = \figwidth]{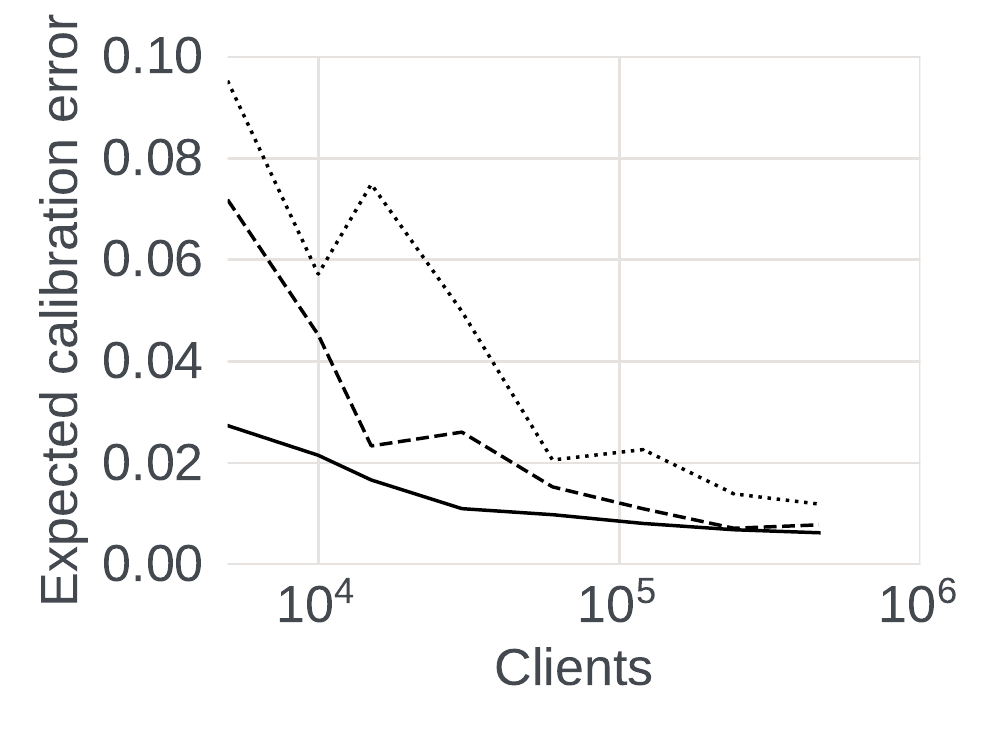}}%
\subcaptionbox{
October 2021 data\label{calib-oct-pop}}{\includegraphics[width = \figwidth]{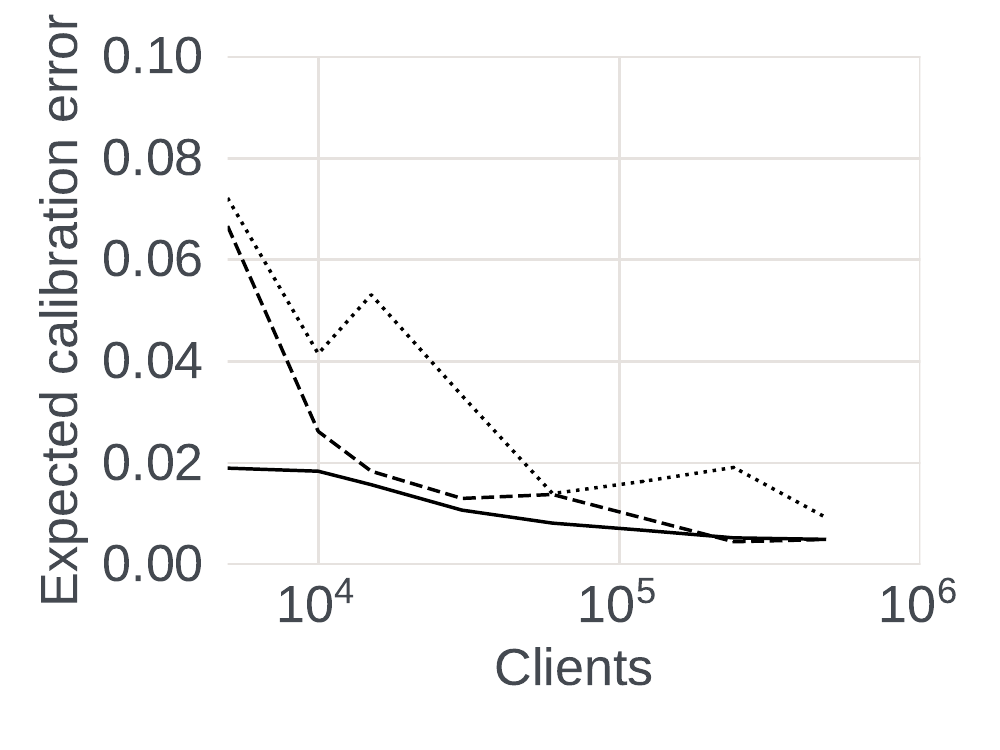}}%
\subcaptionbox{
November 2021 data\label{calib-nov-pop}}{\includegraphics[width = \figwidth]{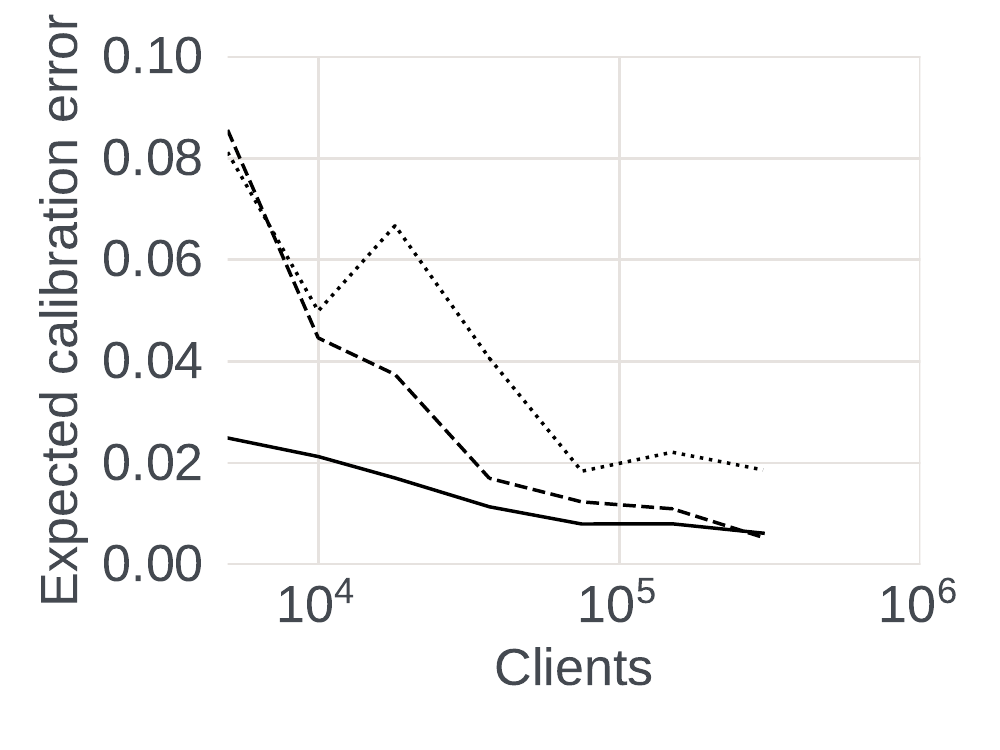}}
\caption{Classifier calibration accuracy with varying population size}
\label{fig:calib-pop}
\end{figure*}

\section{Federated Score Calibration}
\label{sec:calibration}

Classifier calibration poses an additional challenge, since the quality of a calibrated classifier is determined by its performance averaged over multiple examples. When building a summary
{based on a histogram representation of the labeled data}, we incur additional uncertainty: for small buckets holding a few points, our estimates of classifier metrics within such buckets are noisier and subject to sampling error. Hence, \blue{we balance the local precision of smaller buckets with the increased uncertainty} when choosing parameters.
We present results for $(\phi, \ell)$-well-behaved score distributions (Definition~\ref{def:wellbehaved}). 


We start by considering the accuracy in our estimation when we build a score histogram (without privacy noise) with $O(B)$ buckets. 
If the value of the calibrated score function is allowed to vary arbitrarily as the uncalibrated score changes, then calibration via histogram is not a meaningful task. 
As a result, we apply the $(\phi,O(1))$-well-behaved property to the (ideal) calibrated score function for analysis purposes,
ensuring that each bucket is either heavy (contains a spike larger than $\phi = 1/B$), or smooth (obeys the Lipschitz condition with parameter $\ell$).
Hence, we bound the error of the histogram-calibrated function. 

\begin{theoremEnd}{thm}
In the \fed case, the expected calibration error using score histograms is bounded by $O\left(\frac{1}{\clients^{1/3}}\right)$. 
\label{thm:calib:secagg}
\end{theoremEnd}

\begin{proofEnd}
Recall that the (ideal) calibrated value for a score $s$ is the true positive ratio at that point, i.e., $p(s)/(p(s) + n(s))$, where $p(s)$ and $n(s)$ are the \red{probability mass} functions for positive and negative examples.  
Similar to the previous analyses, we will assume that 
the $c(s)$ function is $(1/B, \ell)$-well behaved, for a constant $\ell$, so that between any spikes the maximum change is governed by 
\[|c(s) - c(s + \Delta)| \leq \ell \Delta.\] 

As before, we will make use of score histograms over the positive and negative examples. 
We consider the behavior of the score function within a bucket of width $\Delta$ that includes the score value $s$.
The calibrated value for any point in the bucket must then be in the range $c(s) \pm \ell \Delta = c(s) \pm \ell/B$, since we require that the width of any bucket in the histogram is at most
$1/B$. 

The points drawn in the histogram for this bucket can be considered to be samples, where the probability of each sample for score $s'$ being a positive example is $c(s')$. 
By a standard Hoeffding bound, the probability that the mean calibrated value of $n$ sampled points falls below $c(s) - \ell/B - \varepsilon$, or exceeds $c(s) + \ell/B + \varepsilon$ is bounded by $\exp(-2\varepsilon^2 n)$. 
Since in each bucket we have $\Theta(\clients/B)$ samples, we can set this probability to be a small constant and rearrange to guarantee that for any score $s'$ that falls in the bucket we can estimate its calibrated value within absolute error at most
$\ell/B + \sqrt{B/\clients}$. 
%
%
Otherwise, if the bucket is spiky (contains a spike), then the error is dominated by the sampling error, and so we focus on the non-spiky case. 

Trading off these two error terms, we equate 
$\ell/B = \sqrt{B/\clients}$.
Rearranging, and treating $\ell$ as a constant, we have that
we should set $B = O(\clients^{1/3})$ to balance these errors. 
In this case, the (expected) error achieved is $O(\clients^{-1/3})$. 
\end{proofEnd}

This result shows that calibration is a challenging problem, as accuracy improves slowly as a function of the number of clients, $M$.
This is mostly due to the uncertainty introduced by sampling clients in each bucket.   
Next, we consider the impact of incurring privacy noise. 

\begin{theoremEnd}{thm}
\label{thm:calib:dp}
The expected calibration error in the
\ldp and \cdp cases is given by
$O\left(\frac{1}{\epsilon^{1/2}\clients^{1/4}}\right)$
and
$O\left(\frac{1}{\clients^{1/3}} + \frac{1}{\epsilon\clients^{2/3}}\right)$,
respectively.
\end{theoremEnd}

\begin{proofEnd}
We follow the argument of Theorem~\ref{thm:calib:secagg} to argue that within a bucket we have 
$\Theta(M/B)$ points. 
However, now the estimate from the bucket is perturbed due to privacy noise. 
In particular, we obtain a value for $p_i$ and $n_i$, the numbers of positive and negative examples in the bucket, that have expected absolute error of 
$\sqrt{\clients}/\epsilon$ (in the \ldp case) 
or
$1/\epsilon$ (in the \cdp case). 
In a bucket with $\Theta(\clients/B)$ examples, this yields an additional error on estimates of $c$ of $O(B/\epsilon\sqrt{\clients})$ or $O(B/\epsilon M)$, respectively. 

For the \ldp case, the quantity of $B/\epsilon\sqrt{\clients}$ will dominate the
$\sqrt{B/\clients}$ term, leading us to choose 
$B = O(\sqrt{\epsilon \clients^{1/2}})$. 
This sets the error bound to $O(\epsilon^{-1/2} \clients^{-1/4})$.
That is, we expect the number of bins needed to be rather small in the LDP case. 

For the \cdp case, the quantity of 
$B/\epsilon{\clients}$ will be of lower magnitude than
$\sqrt{B/\clients}$ since (treating $\epsilon$ as a constant) $B/M < 1$. 
Hence, we focus on balancing
$\ell/B$ with $\sqrt{B/\clients}$ as in the noiseless case.
This sets $B = O(M^{1/3})$ to achieve error
$O(\clients^{-1/3} + 1/\epsilon\clients^{2/3})$.
We state this as $O(\clients^{-1/3})$, assuming that 
$1/\clients^{1/3} < \epsilon$. 
\end{proofEnd}

For \cdp, we \red{expect that there are enough clients so that} $1/\clients^{1/3} < \epsilon$, thus the bound simplifies to
$O(\clients^{-1/3})$. 
The dependence on $\epsilon$ is (surprisingly) limited because sampling noise dominates privacy noise. 
Hence, we anticipate small difference in the quality of calibration with and without privacy noise. 
For \ldp, the privacy noise is large enough to affect the error bound but only weakly (as $1/\sqrt{\epsilon}$).

\section{Empirical evaluation}
\label{sec:experiments}

We validated our methods on \red{three different sources of} freely-available data from Kaggle competitions (Table \ref{tab:exp_setup}). 
A small \red{snapshot} of results is shown in Figures~\ref{fig:auc-pop} and \ref{fig:calib-pop} that 
explain how the accuracy of proposed AUC estimation
and classifier calibration vary with population size (\red{number of distributed clients who each hold one labeled example}) for three privacy models. 
 \blue{Notably, we do not assume the use of deep learning and can work with any binary classification model.}
The full experimental results are described in Appendix~\ref{sec:fullexpts}, \red{and confirm our analytical bounds as we now summarize}.

Empirically, the proposed methods for calibration and evaluation work well across \red{diverse data and the range of federated models (\fed, \cdp, \ldp)}. 
For modest parameter settings ($\clients$ and $B$), we observe errors $<0.001$ for all our target metrics. 
This is sufficient for many practical use cases relevant to our study.\eat{applications of federated calibration and evaluation of classifiers.} 
Strengthening the privacy model from \fed to \ldp increases errors, but
the results remain usable across all examples, even with large amounts of noise. Suppose that we want to compare a new classifier to one deployed already, to see if there is a significant improvement in ROC AUC, say, gain $>0.01$ AUC. 
Then even with \ldp noise we can find federated estimates of the AUC with error $<0.005$, sufficient to allow this comparison of classifiers. 
Our experiments also confirm our theoretical analyses and help to choose the key parameters such as the height of the hierarchy $h$ and the number of buckets $B$.  
We observe that in the \fed model, error on AUC estimation is insensitive to the number of clients $\clients$ (see Figure~\ref{fig:auc-pop}), as predicted (see e.g., Table~\ref{tab:results}). 
This error decreases as $\clients$ increases when privacy noise is present, as predicted by Theorems~\ref{thm:auccdp} and~\ref{thm:aucldp}. 
Meanwhile, calibration error decreases in all privacy models as $\clients$ increases (Figure~\ref{fig:calib-pop}). 
Under DP noise, we see good accuracy with population $\clients \approx 10K$-$100K$, comparable to $\clients$ in other \red{descriptions of real-world federated analytics deployments}.

\section{Concluding Remarks}
\label{sec:conclusions}

Full-stack support for {\em federated learning} with classifiers requires {\em federated calibration} and {\em computation of classifier metrics}. Otherwise, leakage of private data after training can make privacy during training moot. Many other aspects of the ML pipeline require similar attention: data cleaning, feature selection and normalization, evaluation of baseline classifiers and and other classifier metrics beyond the ones studied here. 
Some of the components studied here, such as score histograms, can help answering a variety of queries. 
\red{An important feature of our histogram-based approach is that it makes the techniques robust to heterogeneity in the data distributions across clients.}
We saw that privacy can be achieved in the federated and \red{distributed and local} DP models, but increasing accuracy requires participation of more clients as the trust model is narrowed.
We are not aware of any specific negative impacts of this work, beyond the tradeoff between the energy costs associated with federated data analysis and stronger privacy guarantees. 

A natural extension\eat{of this work} \blue{would cover} multi-class classifiers. 
For \red{a few classes} it may suffice to reduce to the binary case (one vs. all) and to build a score histogram for each class. 
\blue{Otherwise, new techniques could} summarize the relationship between the true and predicted labels. 
%
Extending our work to the entire ML pipeline would require \red{some consideration of} {\em privacy budgeting} across tasks to support a single $\epsilon$-DP end-to-end guarantee. 
The challenge is to determine how best to divide $\epsilon$ among different stages and ensure sufficient classifier accuracy. 
The \red{end goal} for this line of work will be to build systems that achieve end-to-end privacy guarantees for federated learning, from feature extraction to deployment with ongoing performance tracking, and so on.
\bibliographystyle{icml2022}
\bibliography{auc}


\newpage
\onecolumn
\appendix

\section{Constructions of score histograms}
\label{app:scorehist}
We formalize the problem of equi-depth histogram as follows. 
Given \clients\ examples with real values $z_i \in[0,1]$ we seek $r_0 \ldots r_B$, so that
\begin{align}
\label{eq:quanthist}
\forall 1 \leq j \leq B: r_{j-1} < r_j,
\text{ and }
| \{ i : r_{j-1} < z_i \leq r_j \}| = \tfrac{\clients}{B}. 
    \end{align}
Note that this definition needs to be revised to handle the case of multiple examples sharing the same value and to tolerate some small approximation, which we gloss over here, to keep the presentation streamlined. 

In the central non-private setting, it is straightforward to find the exact bucket boundaries for a score histogram with equal-weighted buckets: gather all the input data points and sort them, then read off the value after every $\clients/B$ examples. 
This is more challenging for distributed private data,  but has been heavily studied under the banner of finding the quantiles of the input data, so we adopt these approaches~\citep{Cormode:Kulkarni:Srivastava:19,Gaboardi0S19,YangWLCS20}. 
In what follows, we outline finding quantiles under different models of privacy and give the accuracy bounds that result. 
Most federated techniques gather information on the data at a suitably fine granularity, and use this information to find appropriate bucket boundaries.  
For a parameter $h$, we gather information for each integer
$1 \leq k \leq h$, so for $k$
we divide the range $[0,1]$ into $2^k$ uniform segments, each of length $2^{-k}$. 
For each segment, we gather information on how many data points reside in that segment. 
This immediately gives a way to answer a one-dimensional prefix query up to a granularity of $2^{-h}$: given 
a range $[0, r2^{-h}]$ for an integer $r < 2^h$, we can greedily use the computed counts to partition the prefix into at most $h$ segments, at most one for each length $2^{-k}$, for $1 \leq k \leq h$. 
To find the point $q$ such that $|\{ i : z_i < q\}| = \phi$, we can perform a binary search on $r$ to find the best match.

\para{Secure Aggregation.}
The secure aggregation case is most straightforward, since we do not introduce any privacy noise. 
Instead, each client can encode their data into $h$ one-hot vectors of length $2^k$ for $1 \leq k \leq h$. 
This allows the aggregator to find a set of end-points $\hat{r}$ based on~\eqref{eq:quanthist}, 
up to $|\hat{r}_i - r_i| \leq 2^{-h}$. 
When the score distribution is $(\phi,\ell)$-well-behaved, the \red{mass} of client data points varies by at most $\ell$ per unit, 
meaning that the error in the number of points in this approximation is at most $\ell2^{-h}$. 
We can then ensure that the parameter $h$ is chosen so that the error in a bucket, 
$\ell 2^{-h}$, is at most a small fraction of the desired amount (say, $1/4$) of the bucket weight, which is $1/B$. 
Rearranging, we require $h > \log_2 (4B\ell) = 2 + \log_2 B + \log_2 \ell$. 
In other words (treating $\ell$ also as a constant), we only require that $h$ be a constant more than $\log_2 B$. 

\para{\red{Distributed  DP}.}
In the \cdp case, the aggregator obtains the data from the clients 
\red{each of whom add a small amount of noise that collectively adds up to be equal in distribution to some global noise value.
In our setting, we will make use of P\'{o}lya noise which sums to the (discrete) Laplace distribution, a.k.a., the symmetric geometric distribution.}
We can adopt the same depth $h$ hierarchy as before, but now \red{we have discrete Laplace noise added}
to every count (Section~\ref{sec:privacymodels}).  
To ensure differential privacy, the noise has to be scaled as a function of $\epsilon/h$ to achieve an overall $\epsilon$-DP guarantee, since each client contributes to $h$ counts. 
Equivalently, we could divide the clients into $h$ batches, so that each batch reports on a single level of the hierarchy, and adds noise as a function of $\epsilon$. 
In either case, the total variance of finding the number of clients in a range is
$O(h^3/\epsilon^2)$, since $O(h)$ counts each with variance $O(h^2/\epsilon^2)$ are combined.  
This ensures that the end points for bucket boundaries can be found with expected absolute error
$O(h^{3/2}/\epsilon \clients + 2^{-h})$.  
Here, as before, the $2^{-h}$ term comes from representing input points at this granularity.  
If we balance these two terms, then for constant $\epsilon$, and \clients\ between 
$10^3$ and $10^{6}$, then we would expect to choose values of $10\leq h \leq 20$.

\para{Local DP.}
The case of local DP is somewhat similar to the \red{distributed} DP case. 
Now the privacy noise is added by each client independently (typically by a version of randomized response) 
so that their input is encoded in a frequency oracle~\citep{wangetal}. 
Standard approaches apply asymmetric randomized response~\citep{RandResponse1965} to a one-hot encoding of the client's input value, such as the Optimal Unary Encoding approach we adopt in our experiments~\cite{wangetal}. 
In some settings, a client may have no data to report (e.g., we are building a histogram on scores of positive examples, and the client holds a negative example). 
In this case, the client can submit an all zeros vector to the unary encoding mechanism, and obtain the same LDP guarantee for their contribution. 

Each client introduces noise of $O(1/\epsilon^2)$ on each count they report, and
clients are divided into $h$ groups of size $\clients/h$ to report on one level of the hierarchy. 
Now the variance for finding the fraction of clients in a range is
$O(h^2/(\epsilon^2\clients))$, due to the increased amount of noise.
This means that bucket boundary end points are found with expected absolute error
$O(h/\epsilon \sqrt{\clients} + 2^{-h})$. 
This indicates that a shallower hierarchy is preferred: for typical parameter settings, we now expect 
$5\leq h\leq 10$.

\section{Omitted proofs of our results}
\label{app:proofs}
\printProofs

\begin{figure*}[t]
\centering
\includegraphics[trim=875 25 875 25,clip,width=0.4\textwidth]{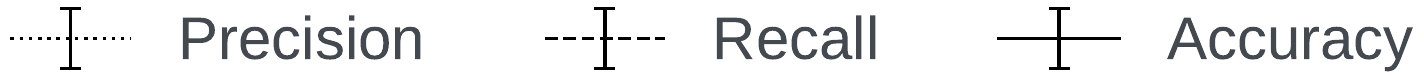}

\hspace*{-9mm}
\subcaptionbox{Sep data, no noise\label{pra-sep-nodp2}}{\includegraphics[width = \figwidth]{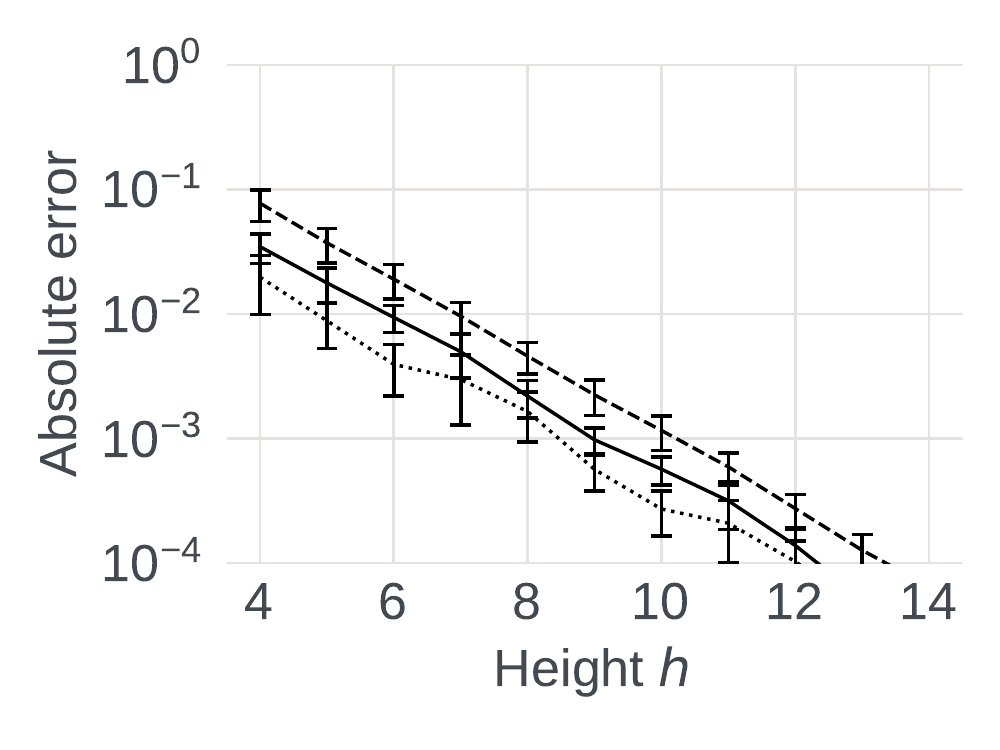}}%
\subcaptionbox{Oct data, no noise\label{pra-oct-nodp2}}{\includegraphics[width = \figwidth]{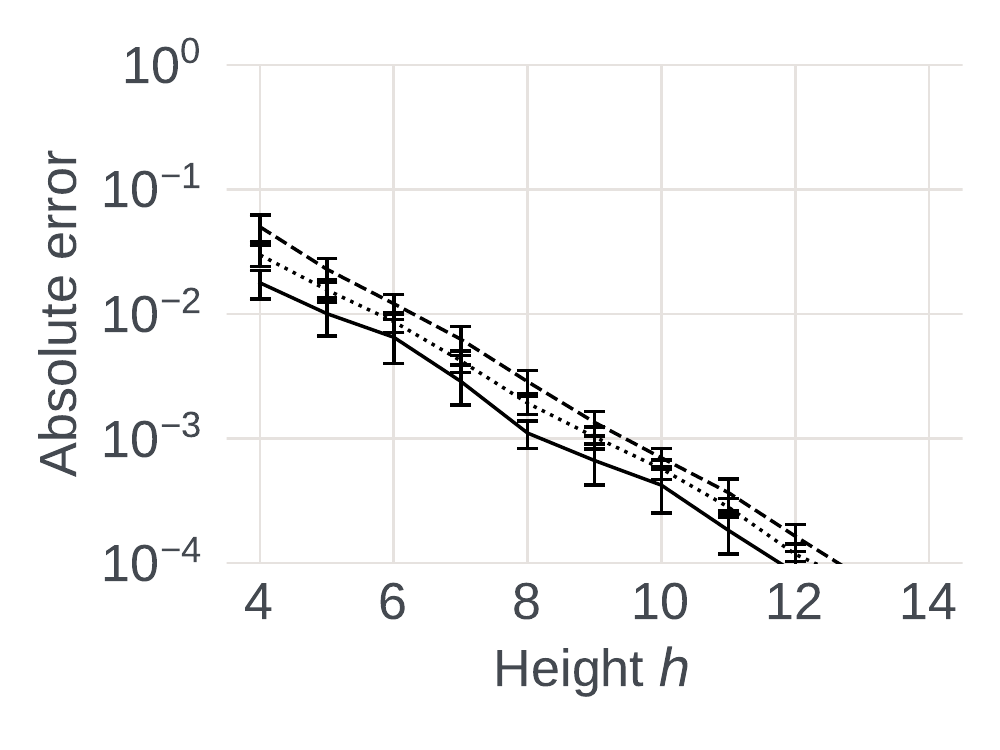}}%
\subcaptionbox{Nov data, no noise\label{pra-nov-nodp2}}{\includegraphics[width = \figwidth]{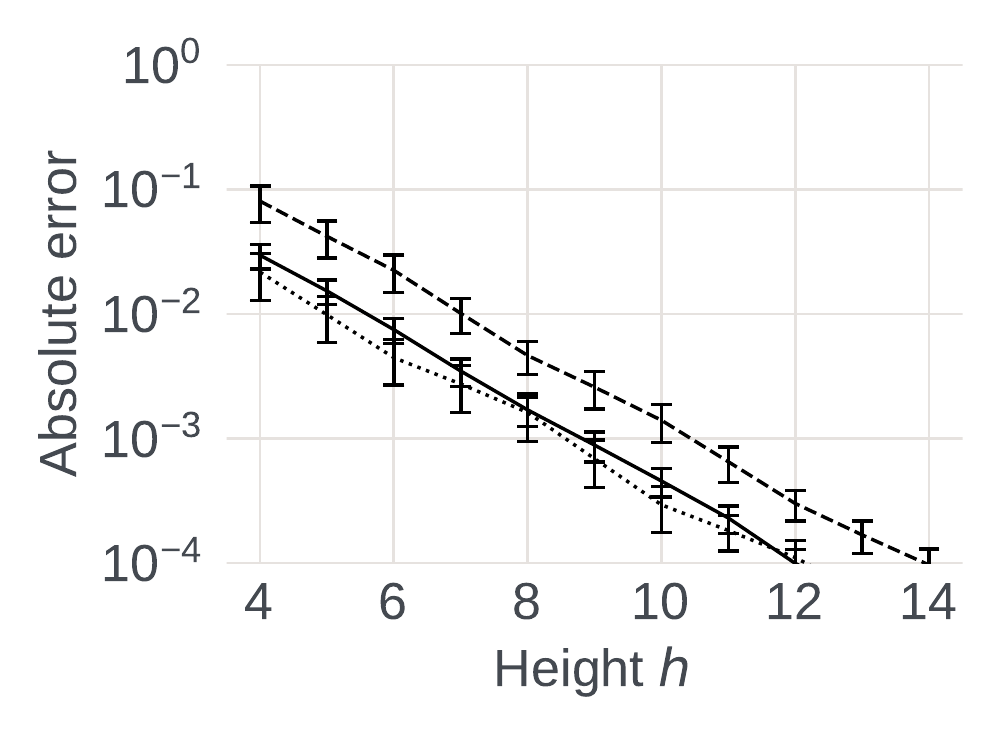}}

\hspace*{-9mm}
\subcaptionbox{Sep data, \cdp noise\label{pra-sep-lap2}}{\includegraphics[width = \figwidth]{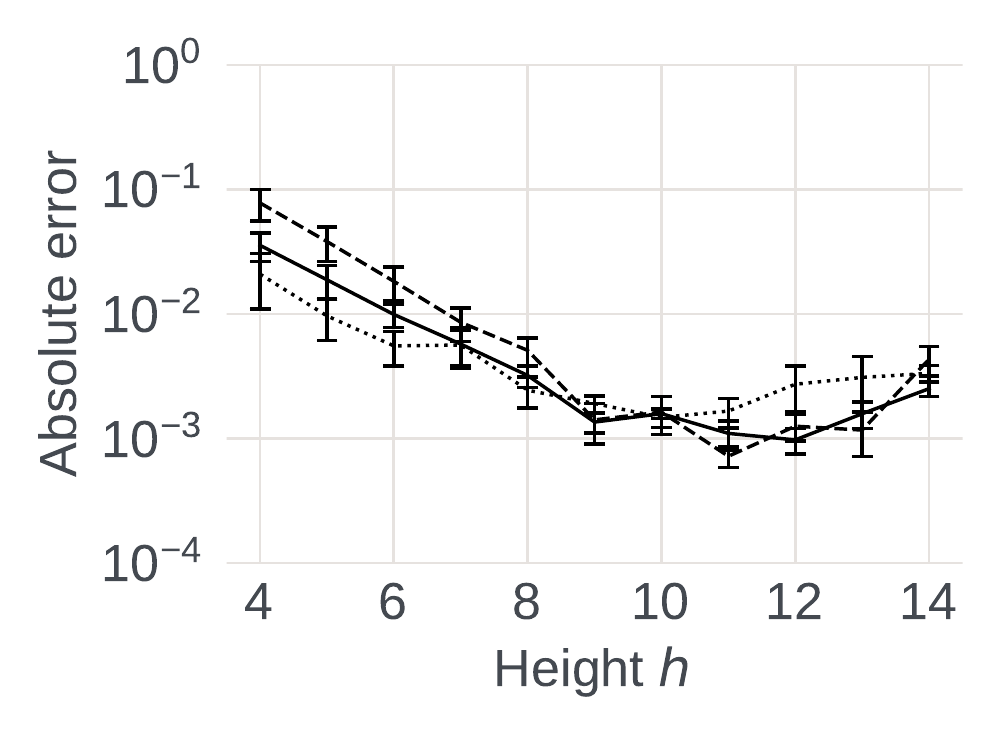}}%
\subcaptionbox{Oct data, \cdp noise\label{pra-oct-lap2}}{\includegraphics[width = \figwidth]{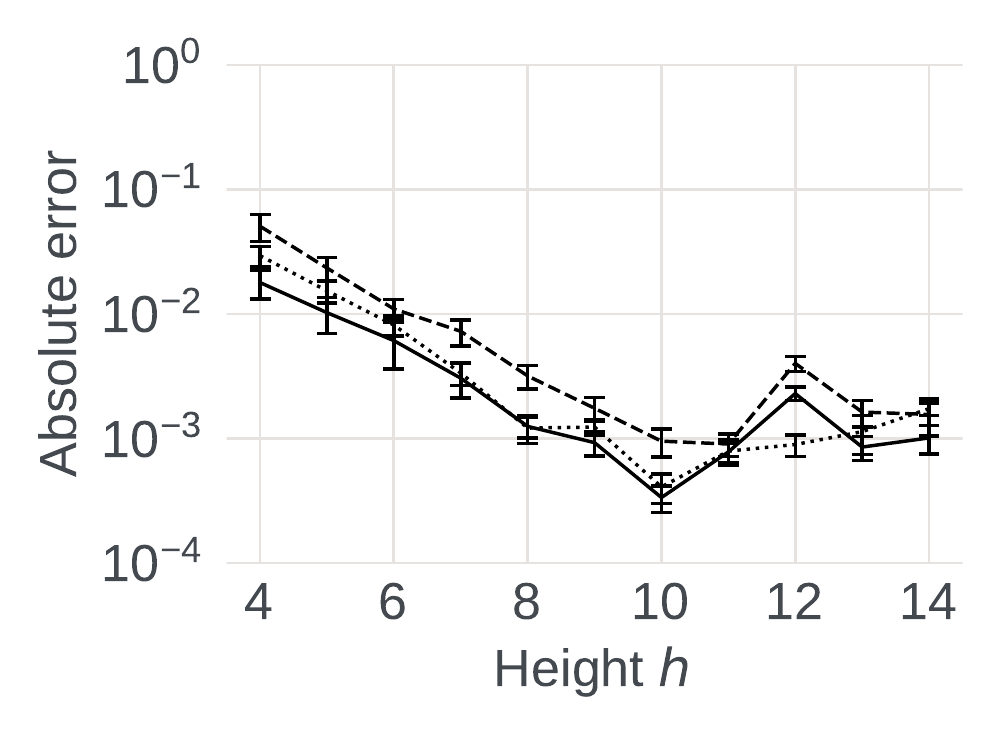}}%
\subcaptionbox{Nov data, \cdp noise\label{pra-nov-lap2}}{\includegraphics[width = \figwidth]{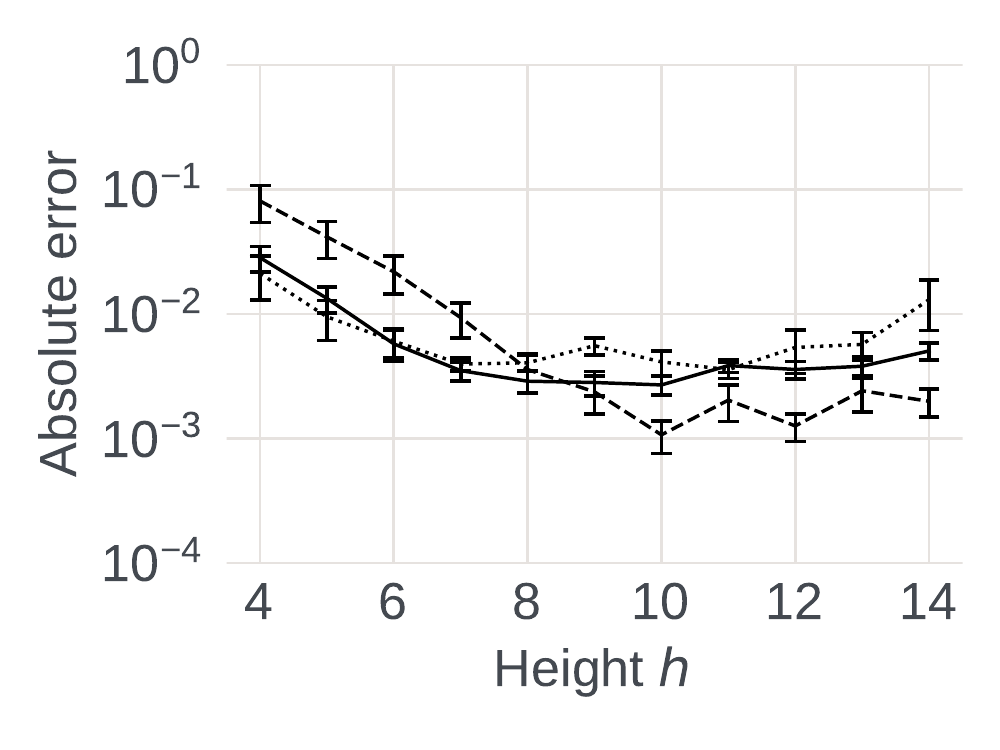}}

\hspace*{-9mm}
\subcaptionbox{Sep data, \ldp noise\label{pra-sep-oue2}}{\includegraphics[width = \figwidth]{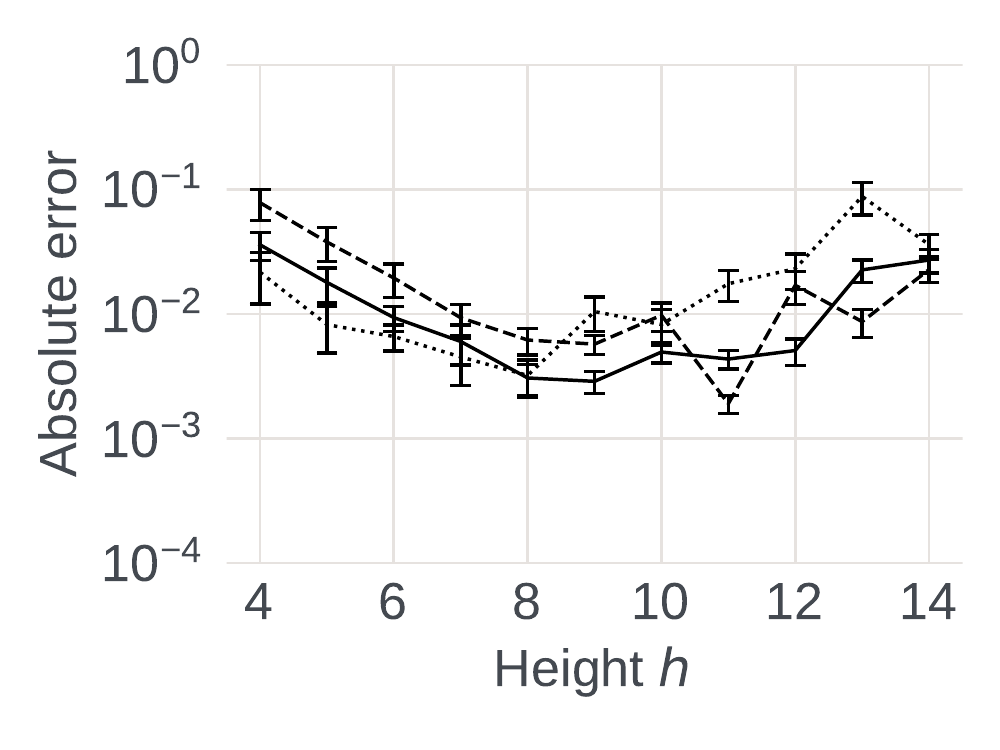}}%
\subcaptionbox{Oct data, \ldp noise\label{pra-oct-oue2}}{\includegraphics[width = \figwidth]{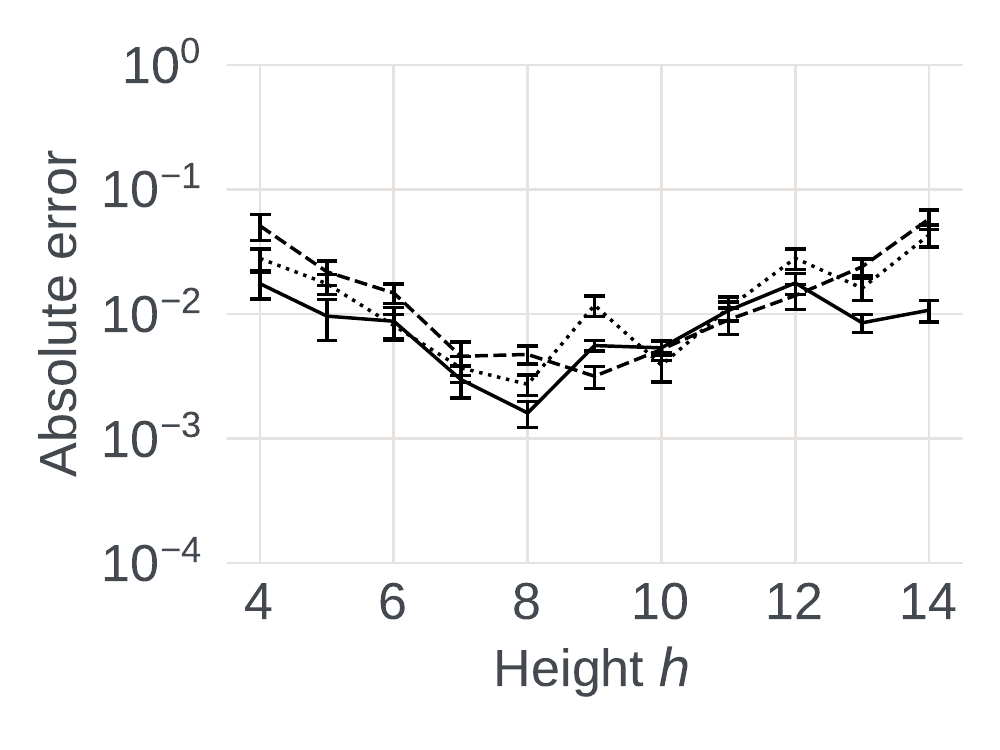}}%
\subcaptionbox{Nov data, \ldp noise\label{pra-nov-oue2}}{\includegraphics[width = \figwidth]{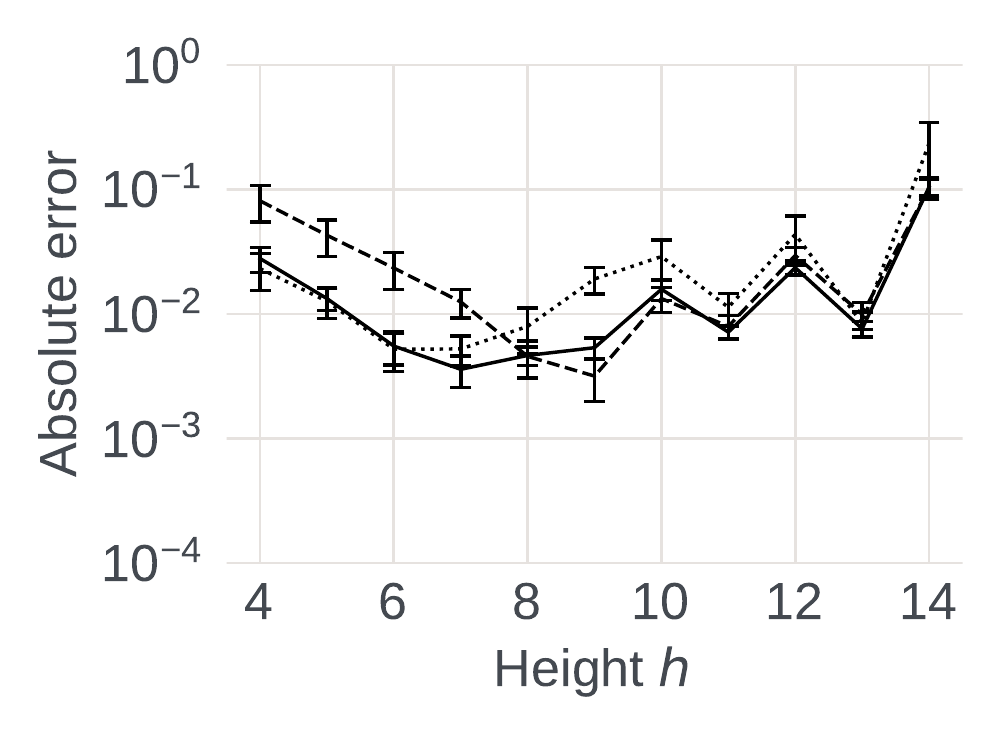}}
\caption{Accuracy for classifier Precision, Recall and Accuracy estimation with varying noise levels}
\label{fig:pra}
\end{figure*}

\begin{figure*}[t]
\centering
\includegraphics[trim=875 25 875 25,clip,width=0.4\textwidth]{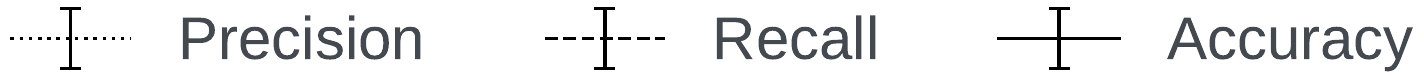}

\hspace*{-9mm}
\subcaptionbox{Sep data, no noise\label{pra-sep-nodp-pop}}{\includegraphics[width = \figwidth]{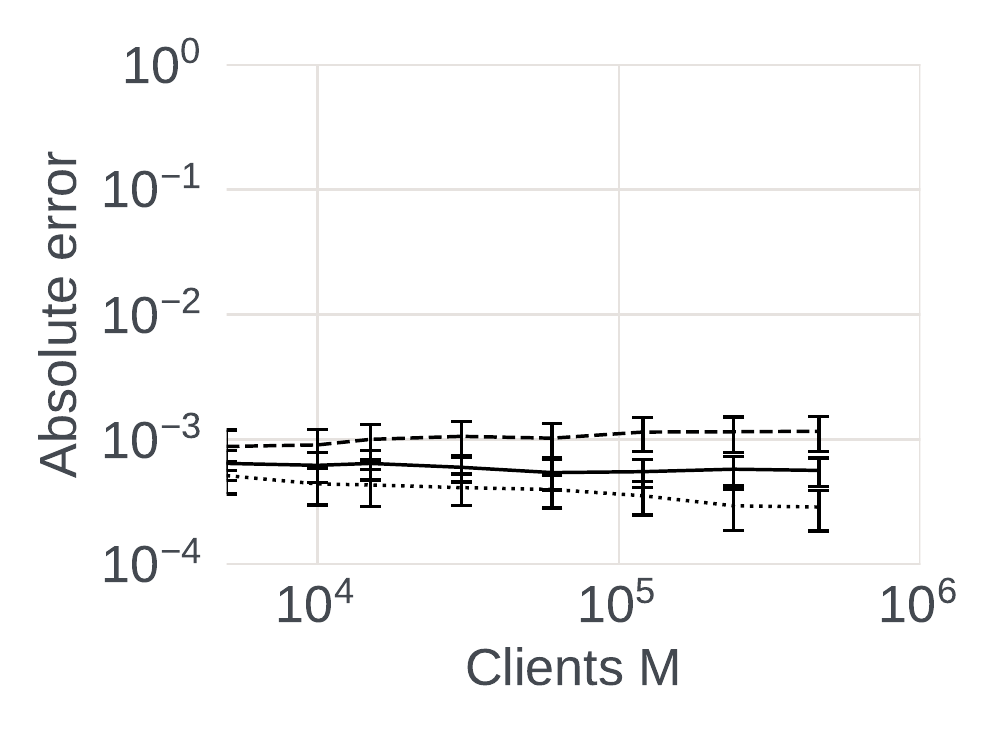}}%
\subcaptionbox{Oct data, no noise\label{pra-oct-nodp-pop}}{\includegraphics[width = \figwidth]{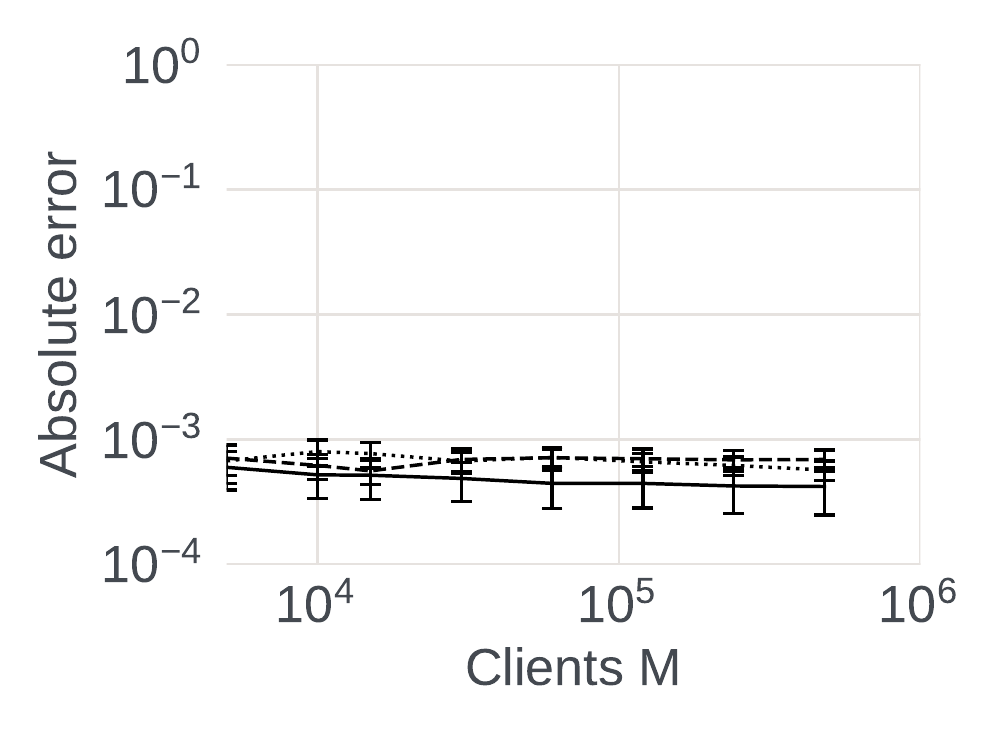}}%
\subcaptionbox{Nov data, no noise\label{pra-nov-nodp-pop}}{\includegraphics[width = \figwidth]{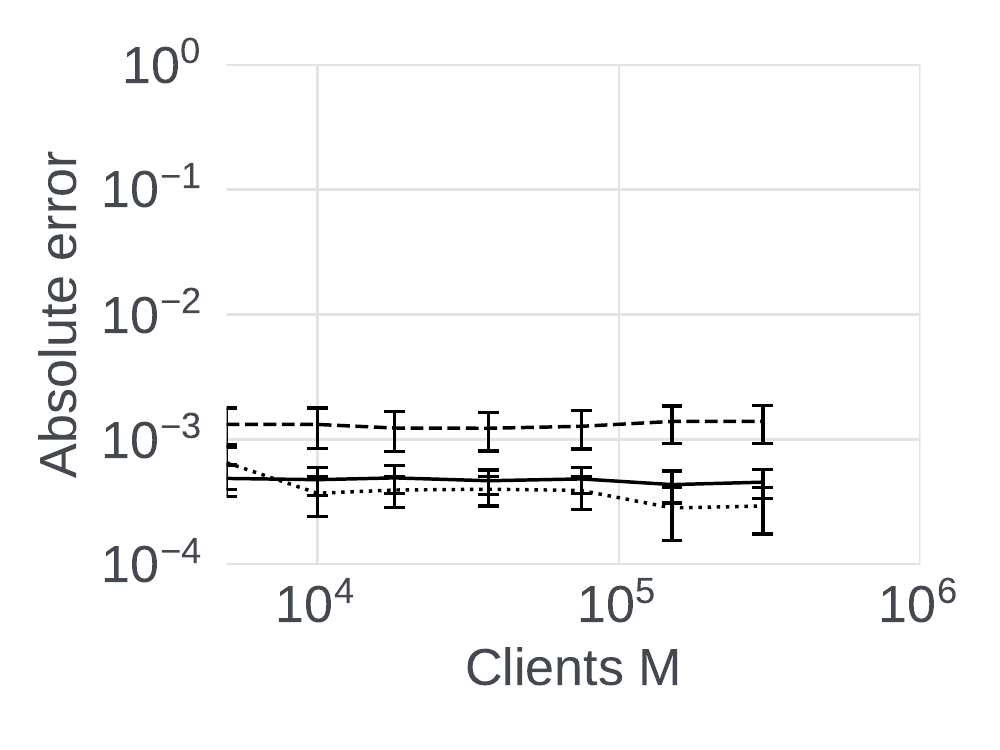}}

\hspace*{-9mm}
\subcaptionbox{Sep data, \cdp noise\label{pra-sep-lap-pop}}{\includegraphics[width = \figwidth]{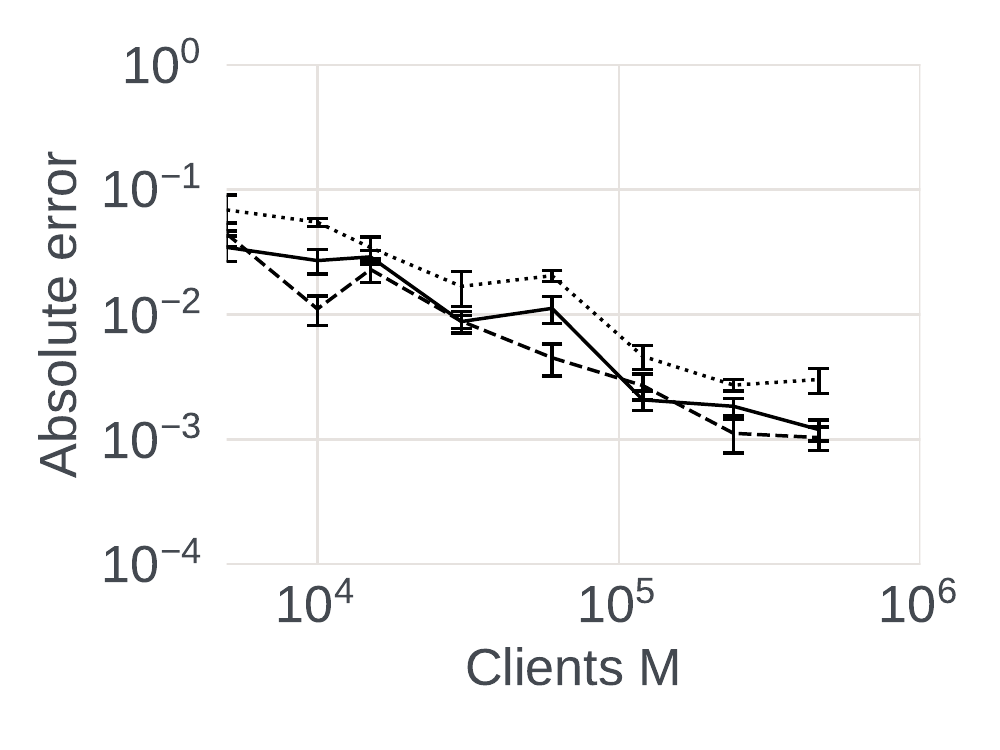}}%
\subcaptionbox{Oct data, \cdp noise\label{pra-oct-lap-pop}}{\includegraphics[width = \figwidth]{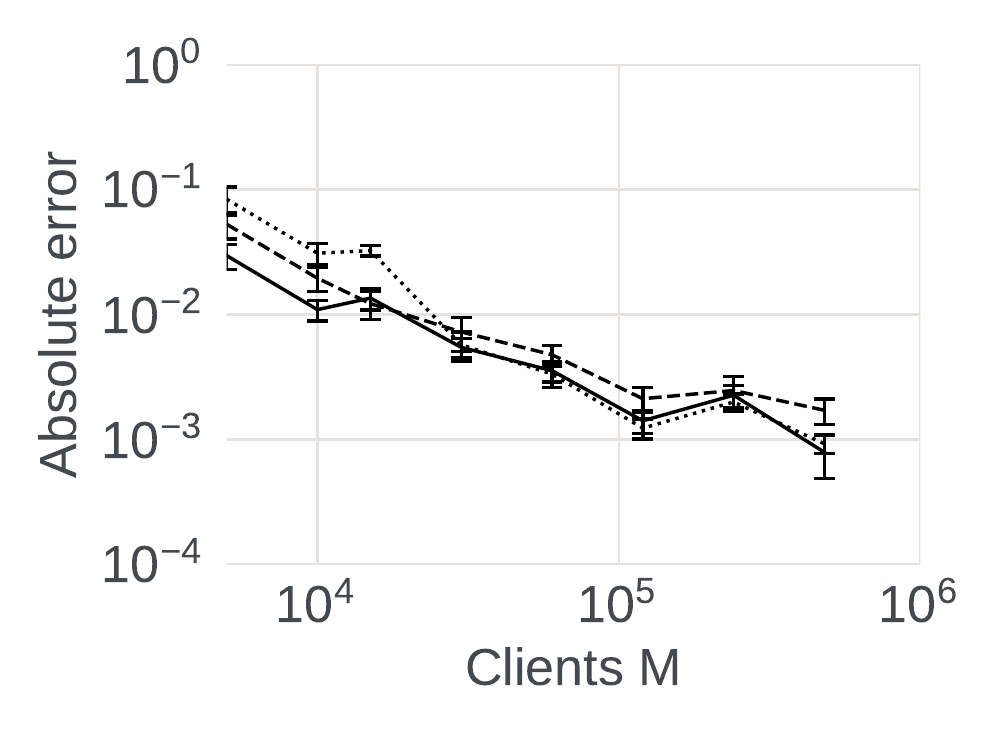}}%
\subcaptionbox{Nov data, \cdp noise\label{pra-nov-lap-pop}}{\includegraphics[width = \figwidth]{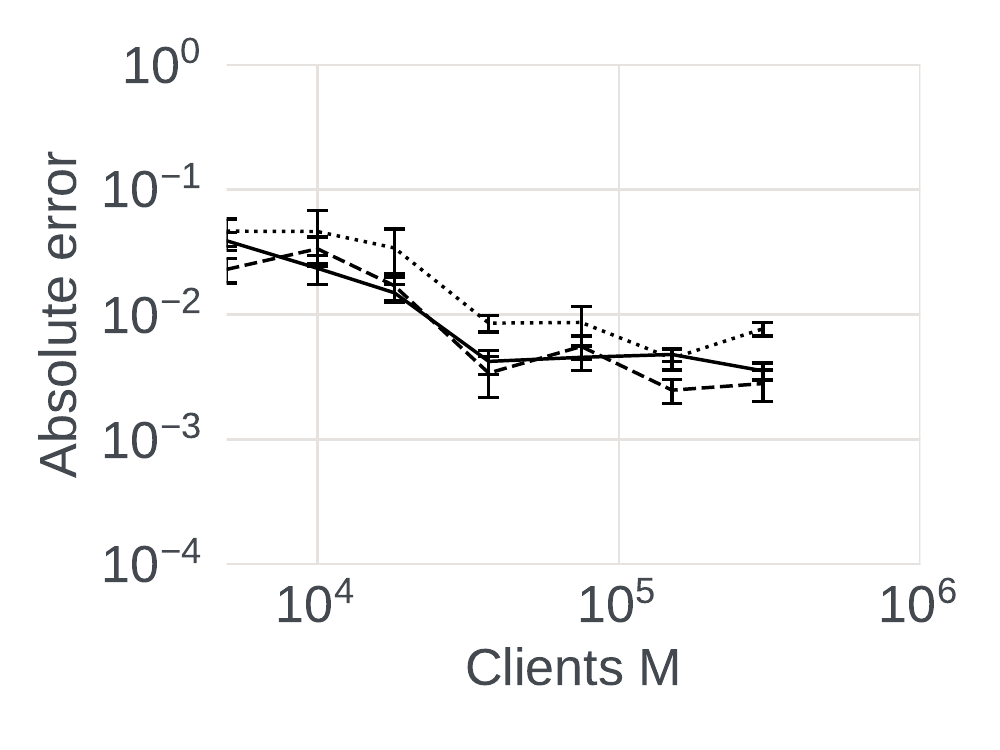}}

\hspace*{-9mm}
\subcaptionbox{Sep data, \ldp noise\label{pra-sep-oue-pop}}{\includegraphics[width = \figwidth]{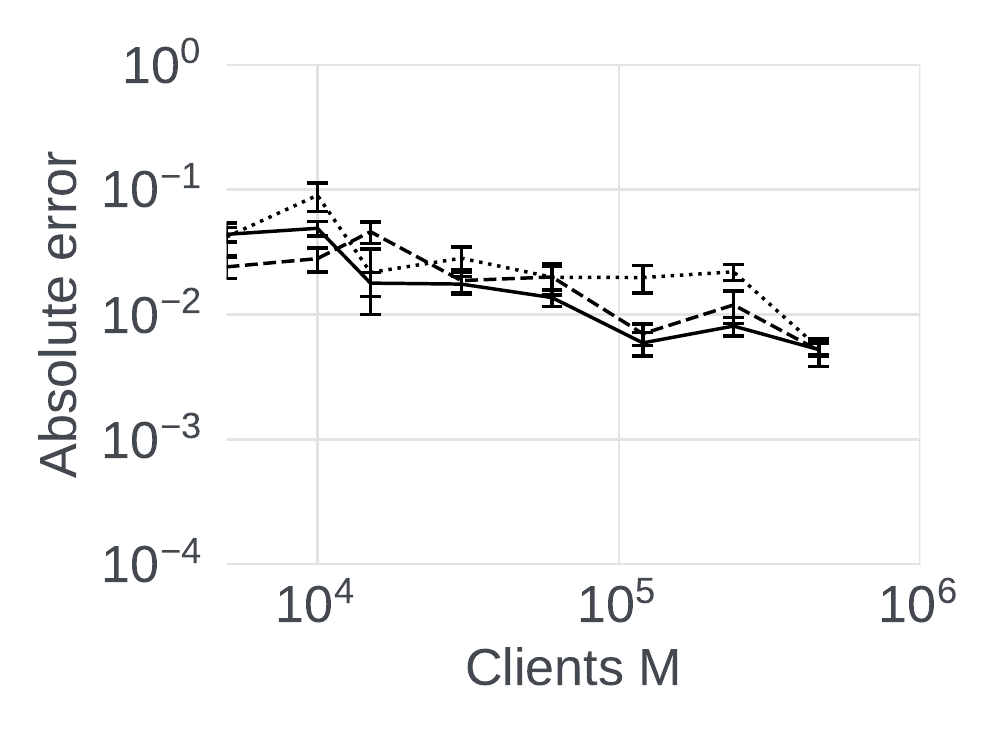}}%
\subcaptionbox{Oct data, \ldp noise\label{pra-oct-oue-pop}}{\includegraphics[width = \figwidth]{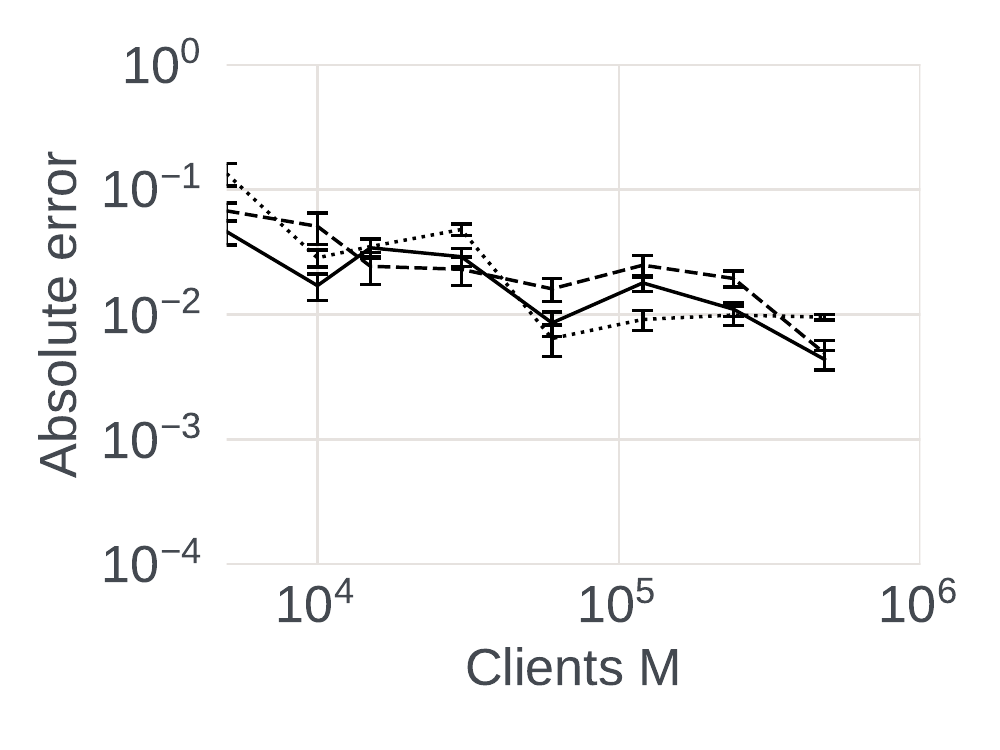}}%
\subcaptionbox{Nov data, \ldp noise\label{pra-nov-oue-pop}}{\includegraphics[width = \figwidth]{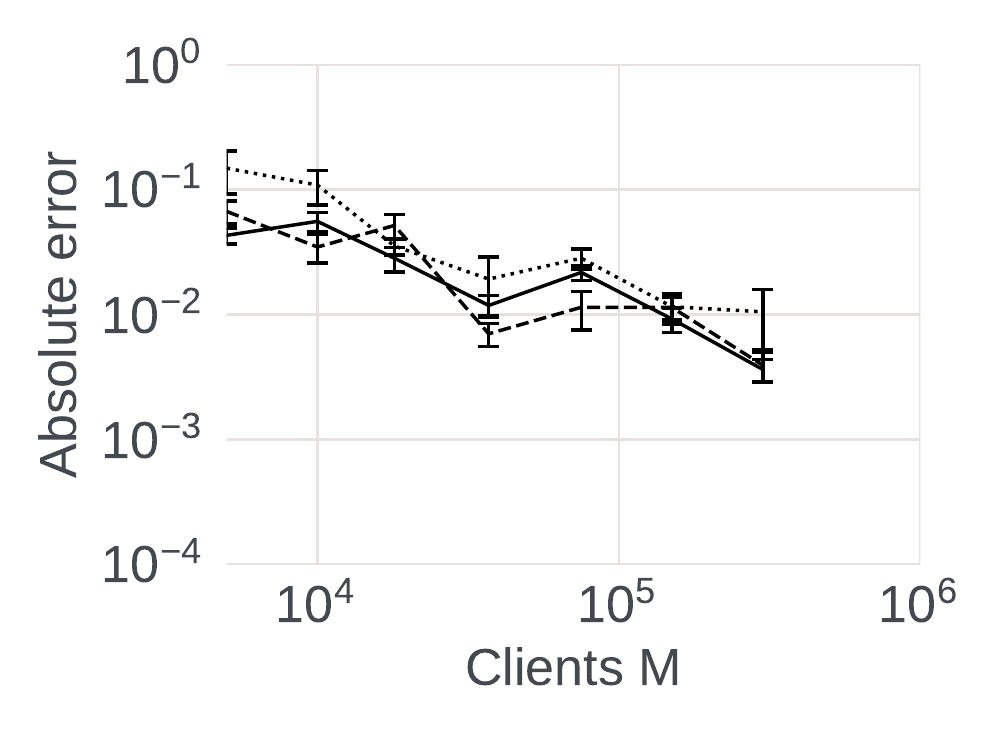}}
\caption{Accuracy for classifier Precision, Recall and Accuracy estimation with varying population size}
\label{fig:pra-pop}
\end{figure*}

\section{Experimental Evaluation}
\label{app:expts}
\label{sec:fullexpts}

Our empirical study quantifies the accuracy of the proposed approaches and the impact of enforcing different models of privacy. To this end, our Python notebook (provided as supplementary material) ($i)$ simulates a distributed environment on a single CPU and ($ii$) evaluates several approaches for a selection of trained classifiers using data examples with ground-truth labels and predicted scores (the nature of the features is unimportant here).

We leverage freely-available synthetic data and trained baseline classifiers from \red{three} recent ``tabular playground'' competitions at Kaggle.\footnote{See \url{https://www.kaggle.com/competitions?hostSegmentIdFilter=8}}\eat{The Kaggle data science platform hosts a number of competitions.}
These data science challenges present realistic synthetic data and prediction objectives. 
They are more difficult than introductory-level challenges and amenable to a range of approaches.
Table \ref{tab:exp_setup} summarizes 
the data and classifiers we used from three \red{different} Kaggle challenges 
\red{from September 2021 (Sep), October 2021 (Oct) and November 2021 (Nov). 
They each have} binary targets and approximately balanced positive and negative classes.\eat{
Each challenge offers up to 1M labeled examples (958K for Sep, 1M for Oct and 600K for Nov), 
We implemented baseline classifiers using code provided by Kaggle users -- a LightGBM classifier (Sep), an XGBoost classifier (Oct) and logistic regression (Nov). 
These classifiers attain respectable ROC AUC scores (0.79 for Sep, 0.85 for Oct and 0.73 for Nov), even though our work does not assume having competitive classifiers.
}
We report accuracy results for these three classifiers in the context of local differential privacy (\ldp), \red{distributed} differential privacy (\cdp), and no noise addition. For \ldp, we make use of Optimized Unary Encoding~\citep{wangetal} with $\epsilon=5.0$, a common setting, which was slightly preferable to other \ldp mechanisms. 
For \cdp, we \red{generate discrete Laplace noise via the summation of $\clients$ P\'{o}lya distributions}, equivalent to $\epsilon = 1.0$. 
In each experiment, we split the data between train and test evenly.

\begin{figure*}[th]
\centering
\includegraphics[trim=600 25 600 25,clip,width=1.0\textwidth]{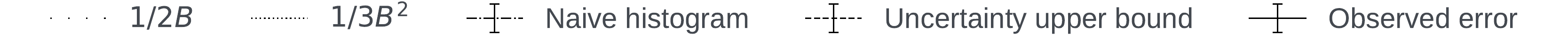}\newline
\hspace*{-9mm}
\subcaptionbox{Sep data, no noise\label{auc-sep-nodp2}}{\includegraphics[width = \figwidth]{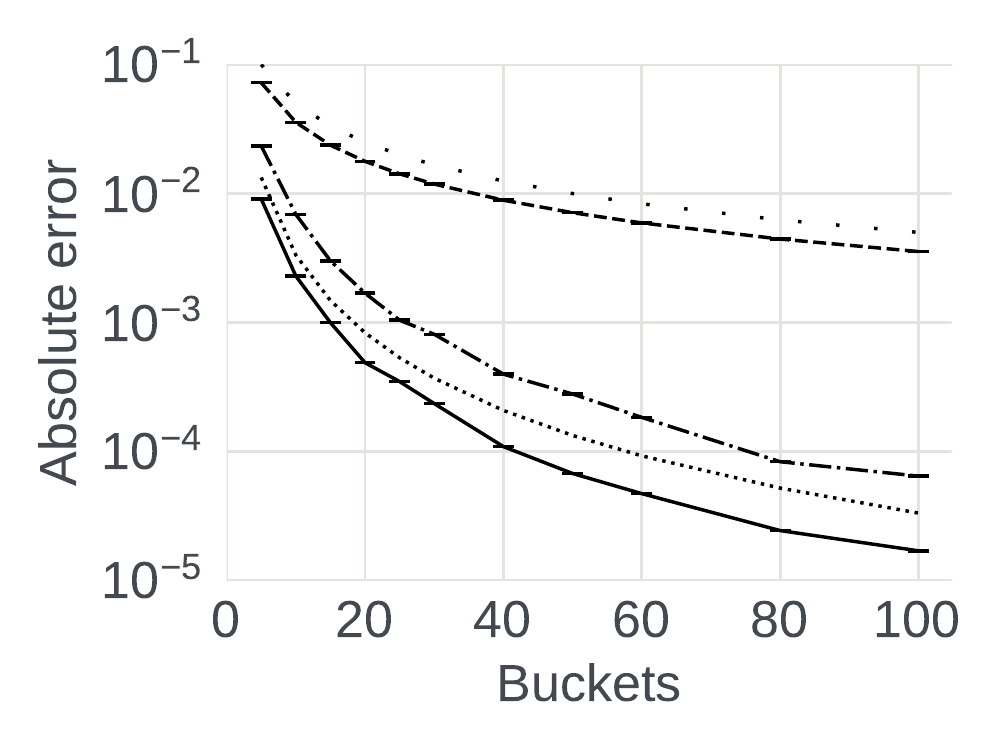}}%
\subcaptionbox{Oct data, no noise\label{auc-oct-nodp2}}{\includegraphics[width = \figwidth]{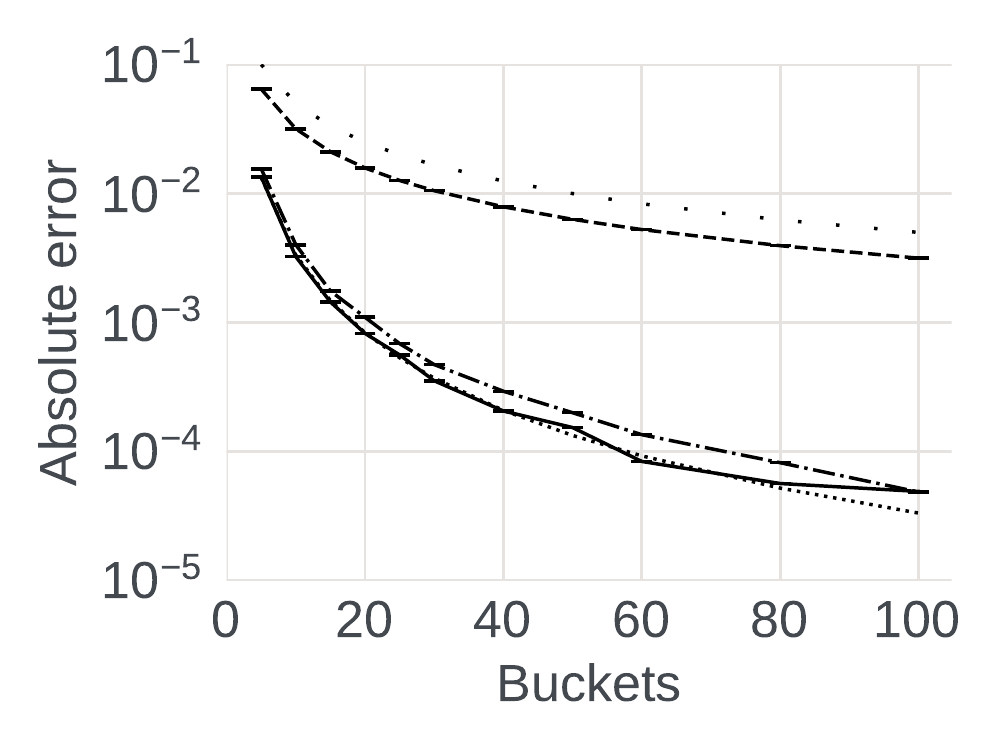}}%
\subcaptionbox{Nov data, no noise\label{auc-nov-nodp2}}{\includegraphics[width = \figwidth]{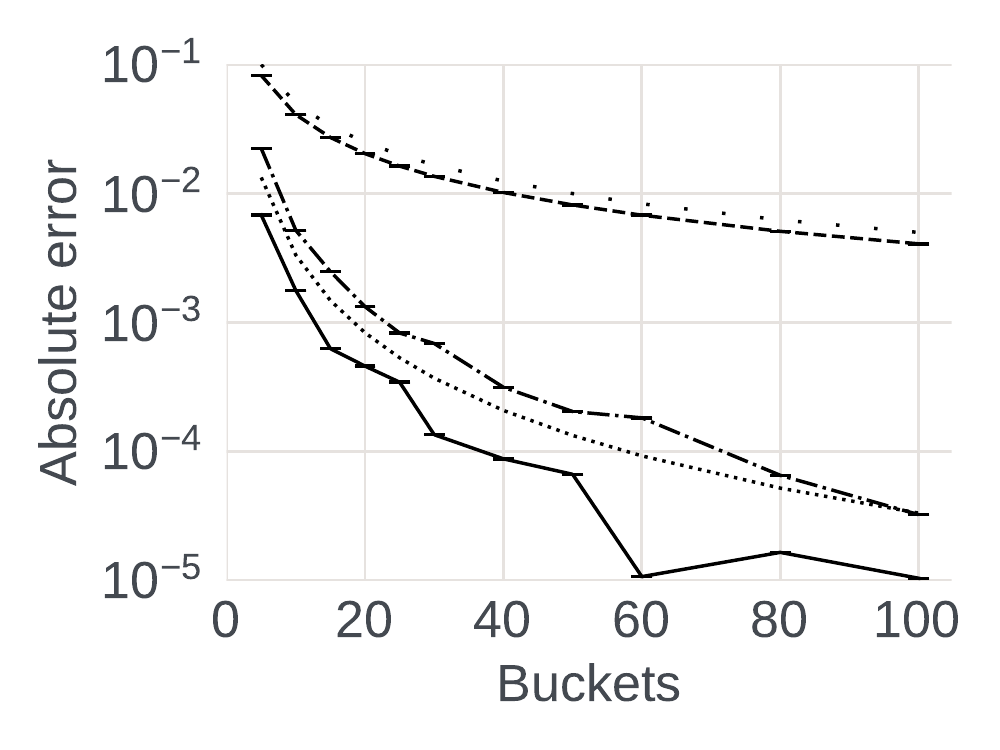}}

\hspace*{-9mm}
\subcaptionbox{Sep data, \cdp noise\label{auc-sep-lap2}}{\includegraphics[width = \figwidth]{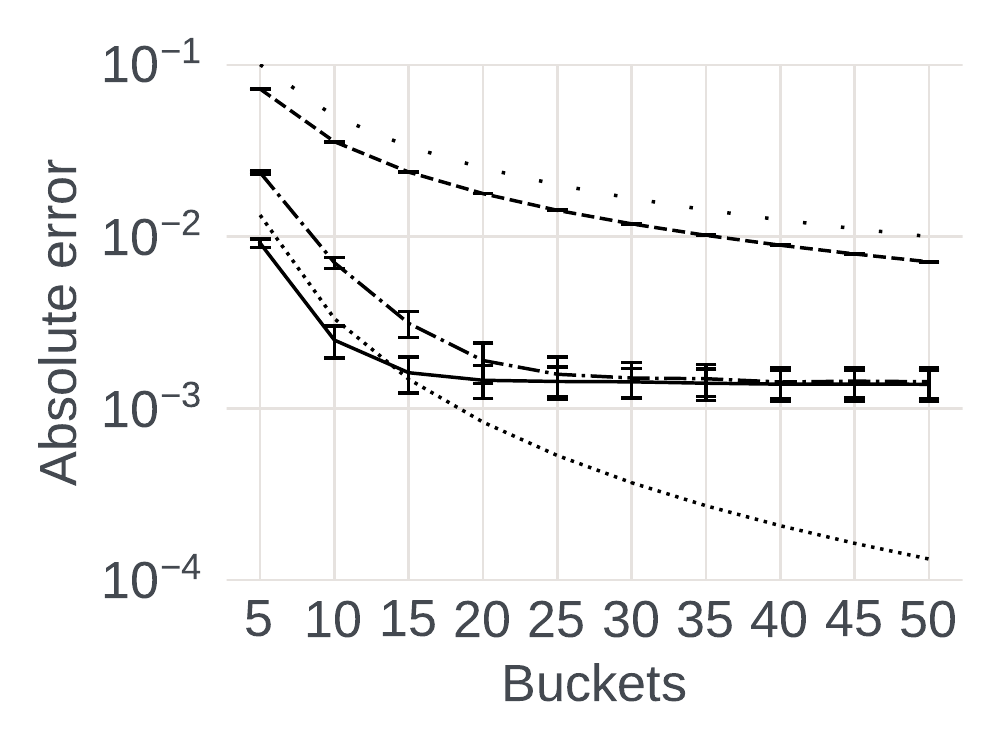}}%
\subcaptionbox{Oct data, \cdp noise\label{auc-oct-lap2}}{\includegraphics[width = \figwidth]{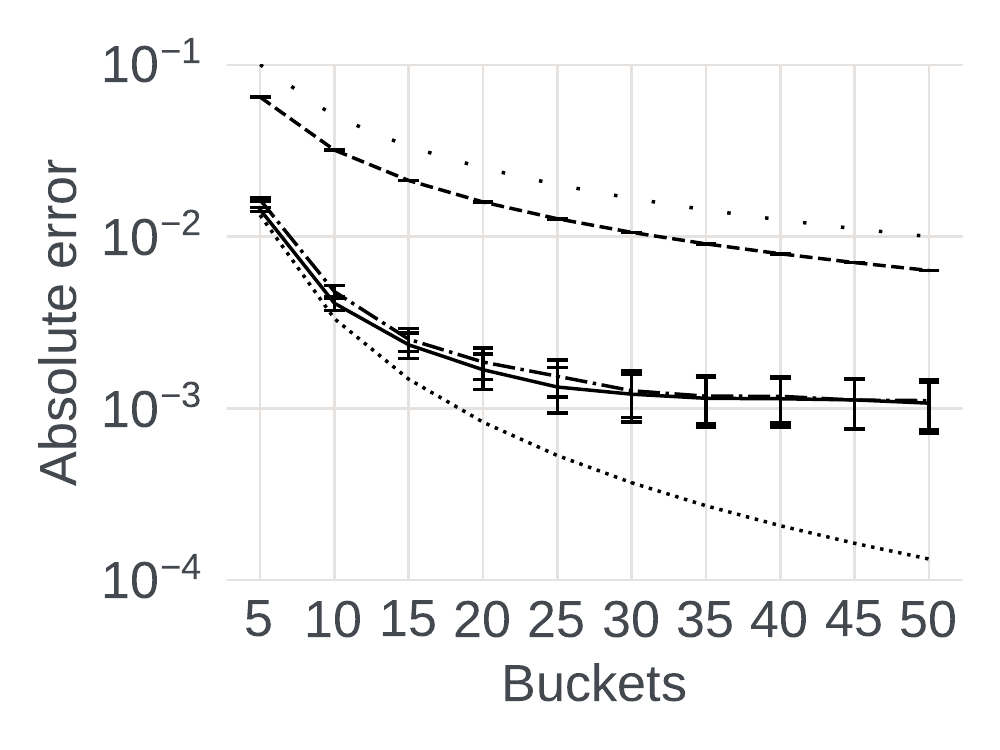}}%
\subcaptionbox{Nov data, \cdp noise\label{auc-nov-lap2}}{\includegraphics[width = \figwidth]{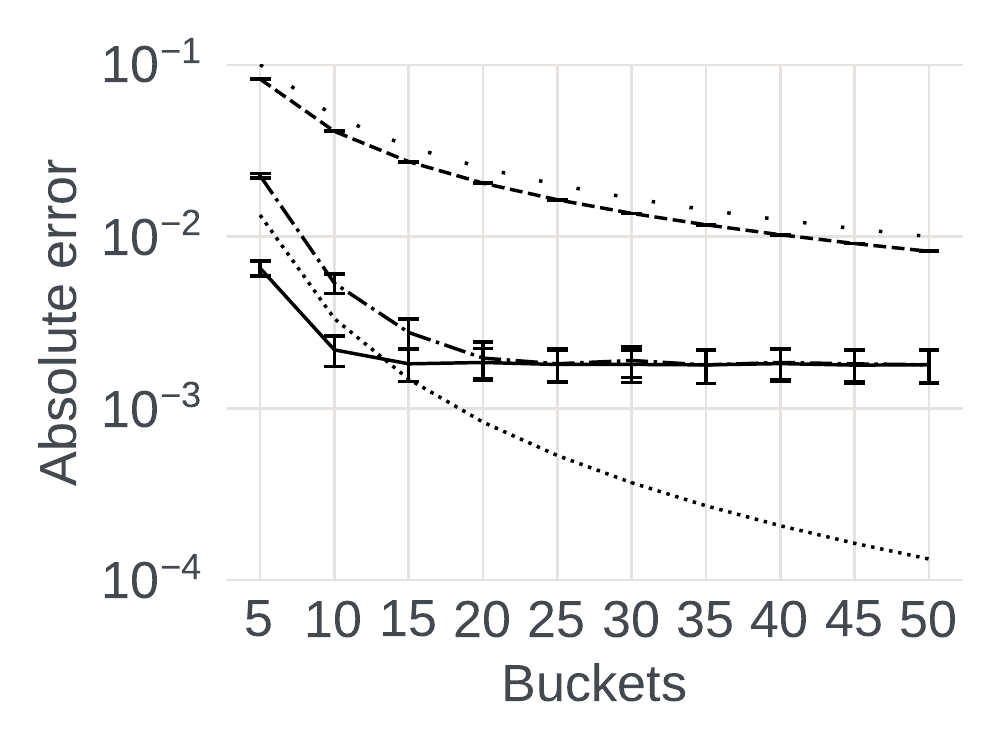}}

\hspace*{-9mm}
\subcaptionbox{Sep data, \ldp noise\label{auc-sep-oue2}}{\includegraphics[width = \figwidth]{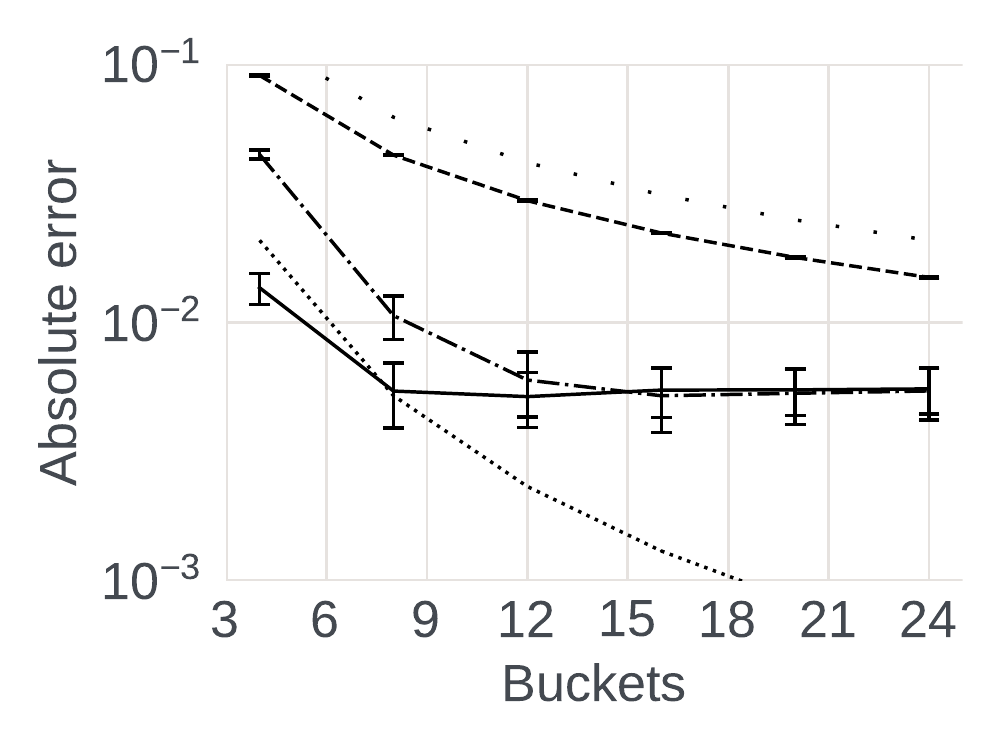}}%
\subcaptionbox{Oct data, \ldp noise\label{auc-oct-oue2}}{\includegraphics[width = \figwidth]{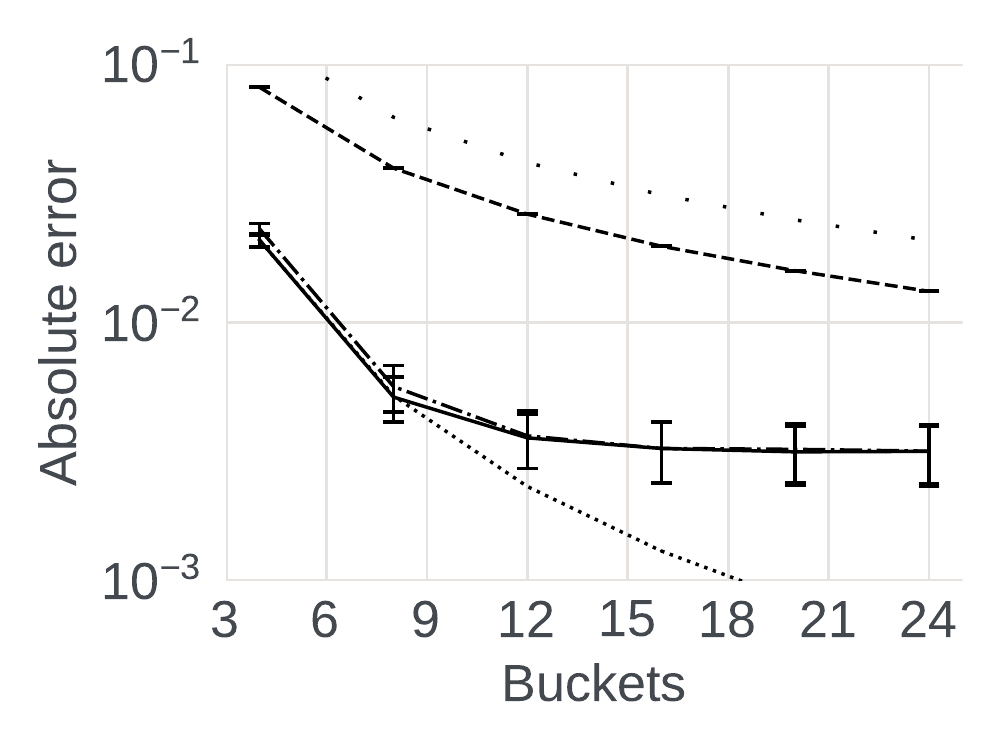}}%
\subcaptionbox{Nov data, \ldp noise\label{auc-nov-oue2}}{\includegraphics[width = \figwidth]{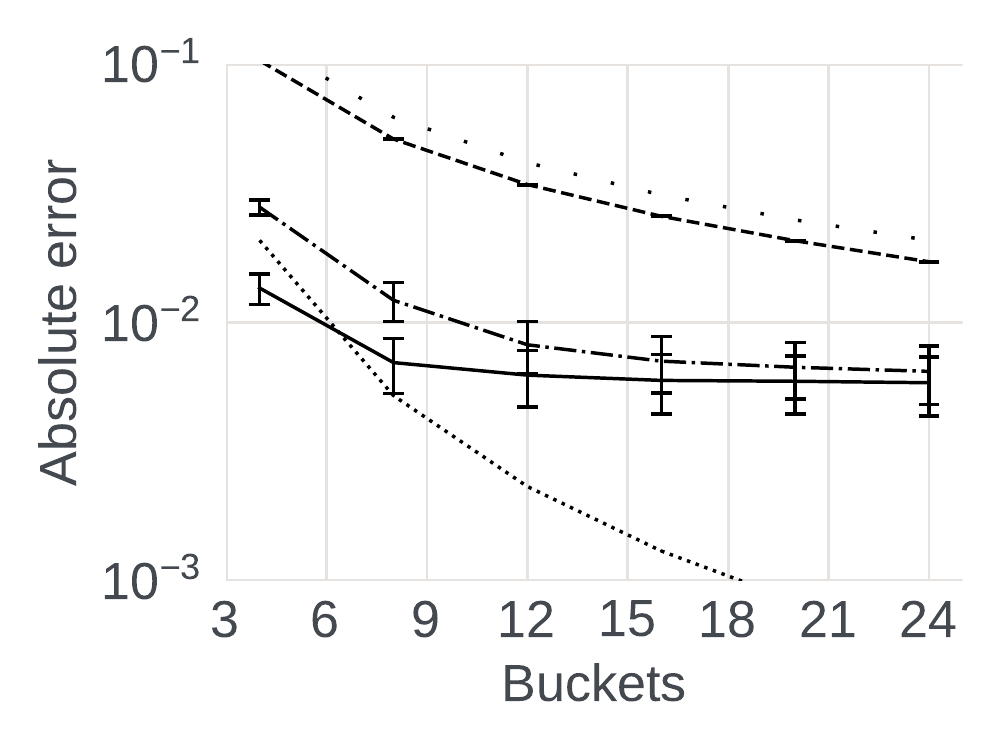}}
\caption{Accuracy for ROC AUC estimation with varying noise levels}
\label{fig:auc}
\end{figure*}

\eat{
\begin{figure*}[th]
\centering
\includegraphics[trim=800 25 800 25,clip,width=0.6\textwidth]{figs/auc_pop_legend.pdf}\newline
\subcaptionbox{Sep data\label{auc-sep-pop}}{\includegraphics[width = \figwidth]{figs/auc-sep-pop.pdf}}%
\subcaptionbox{Oct data\label{auc-oct-pop}}{\includegraphics[width = \figwidth]{figs/auc-oct-pop.pdf}}%
\subcaptionbox{Nov data\label{auc-nov-pop}}{\includegraphics[width = \figwidth]{figs/auc-nov-pop.pdf}}
\caption{Accuracy for ROC AUC estimation with varying population size}
\label{fig:auc-pop}
\end{figure*}
}

\subsection{Precision, Recall, and Accuracy}
\label{app:praexpts}
Figure~\ref{fig:pra} shows results for estimating precision, recall and accuracy over the three data sets, as we vary the parameter $h$ that determines the height of the hierarchy used to build the score histogram. 
For each data set, we consider ten different decision score thresholds ($1/11, 2/11 \ldots 10/11$) to define a binary classifier. 
For each experiment, we show the absolute error between the exact and reported values, with error bars showing the variation over the ten threshold choices.  
Figures~\ref{pra-sep-nodp2}-\ref{pra-nov-nodp2} show that 
in the \fed setting the error behavior
is similar across three datasets. 
Error decreases rapidly as the height is increased, since the only error is due to the approximation induced by the number of cells in the histogram, which are able to better capture the score distribution with increasing height (and hence number of cells). Theorem~\ref{thm:pra} predicts that error decreases proportionally to $1/B = 2^{-h}$, and we see this in practice: a decrease by an order of magnitude when the height $h$ increases by just over 3. 
Of the three metrics, {\em precision} has slightly higher error, consistent with the observations in the proof of Theorem~\ref{thm:pra}: precision depends on the ratio between the number of examples that are correctly classified as positive and the total number of examples that are classified as positive, both quantities estimated using the histogram. In contrast, for {\em recall} and {\em accuracy}, we only have uncertainty in the numerator of the ratio. 
The total error can be made arbitrarily small, e.g., $<10^{-4}$ for $h=14$, which is sufficiently small to compare two classifiers correctly. 

Introducing DP noise to the histograms (Figures~\ref{pra-sep-lap2}-\ref{pra-nov-lap2}) lower-bounds the error.
As predicted by Theorem~\ref{thm:pra}, there is now a tradeoff between ($i$) better data descriptions with a taller hierarchy, and ($ii$) the extra privacy noise due to more numerous buckets. The analysis in the proof of Theorem~\ref{thm:pra} suggests setting the number of buckets $B$ proportional to $(\epsilon M)^{2/3}$.  
In these experiments with $\epsilon = 1.0$ and $M$ in the range of hundreds of thousands, we should choose the height close to 12. Indeed, we observe the lowest error near this predicted value, near $h=11$ for Sep, $h=10$ for Oct, and $h=10$ for Nov. 
These yield absolute error values around 0.001, which is still small enough to compare alternate classifiers. 

For \ldp noise (Figures~\ref{pra-sep-oue2}-\ref{pra-nov-oue2}), the tradeoff is shallower, and the lowest error is seen around 0.005. 
This is large enough to complicate classifier comparisons, but still small enough to track the performance of a deployed classifier and raise alerts when precision, recall or accuracy deviate from historical values.
The analysis in Theorem~\ref{thm:pra} suggests choosing the number of buckets proportional to $\epsilon^{2/3} M^{1/3}$, which means $h=8$ for these experiments and indeed marks the lowest error for estimating precision, recall and accuracy.

Our analysis assumes a hierarchy of height $h$ with fanout 2 and $B = 2^h$ buckets. The analysis generalizes to other fanouts $f$ with $B = f^h$. In additional experiments
with $f=3,4$, the observed error behavior was similar as a function of $B$, i.e., small changes in fanout $f$ did not lead to reduction in error. 

In Figure~\ref{fig:pra-pop}, we study the impact of varying the number of clients, $\clients$ across the three privacy regimes. Throughout, we set $h=10$ based on the previous experiments. Now there is no observable impact in the \fed case (Figures~\ref{pra-sep-nodp-pop} to~\ref{pra-nov-nodp-pop}): increasing $\clients$ does not affect the accuracy. 
For \cdp noise (Figures~\ref{pra-sep-lap-pop} to~\ref{pra-nov-lap-pop}) 
and \ldp noise (Figures~\ref{pra-sep-oue-pop} to~\ref{pra-nov-oue-pop}), 
errors drop as $\clients$ increases, consistent with the $O(1/M)$ behavior noted in Theorem~\ref{thm:pra}. 
That is, increasing $\clients$ by 10x reduces error by as much. Getting good accuracy with DP noise requires a population $>10$K, and $>100$K for \ldp. 

\begin{figure*}[t]
\centering
\includegraphics[trim=600 25 600 25,clip,width=0.8\textwidth]{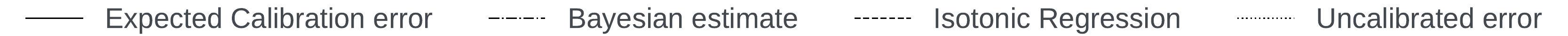}\newline
\hspace*{-9mm}
\subcaptionbox{Sep data, no noise\label{calib-sep-nodp2}}{\includegraphics[width = \figwidth]{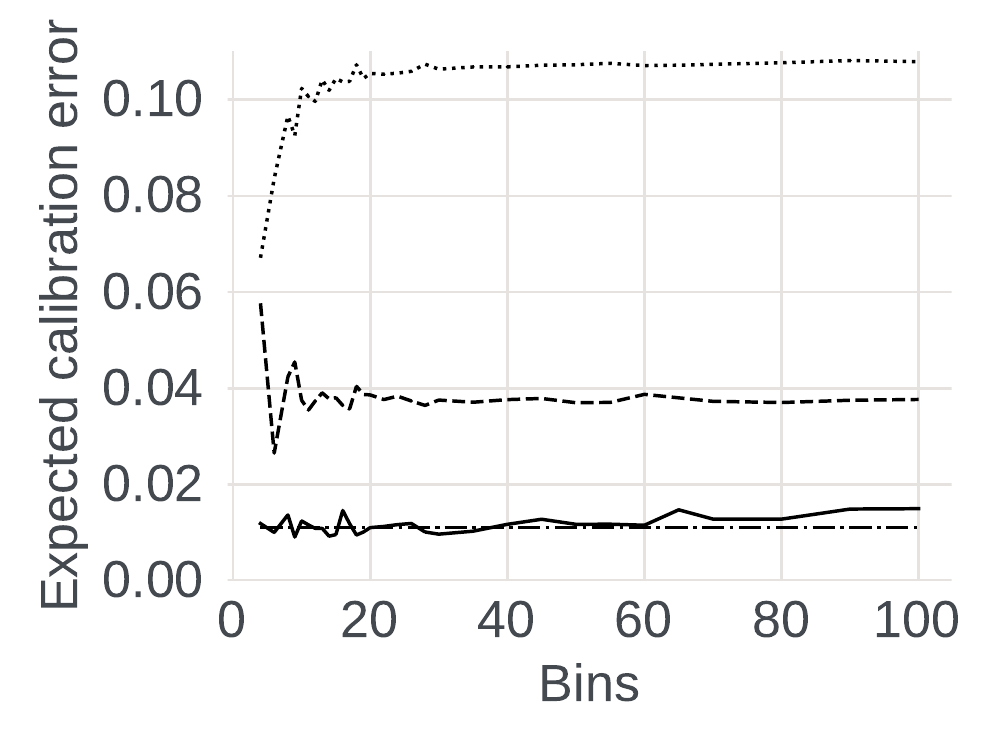}}%
\subcaptionbox{Oct data, no noise\label{calib-oct-nodp2}}{\includegraphics[width = \figwidth]{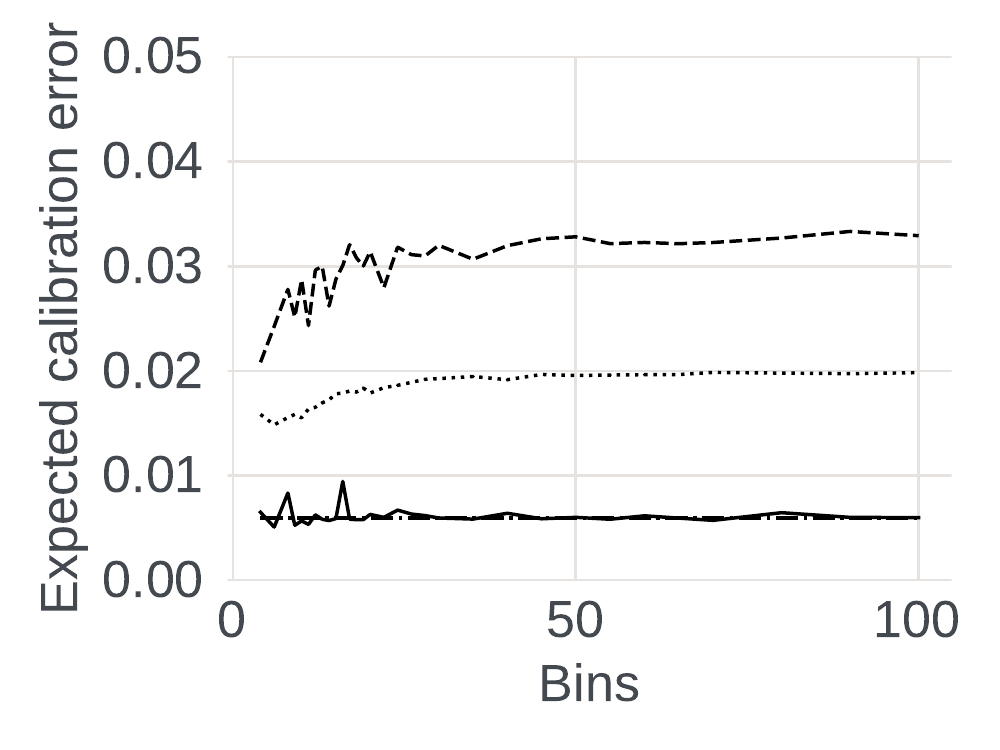}}%
\subcaptionbox{Nov data, no noise\label{calib-nov-nodp2}}{\includegraphics[width = \figwidth]{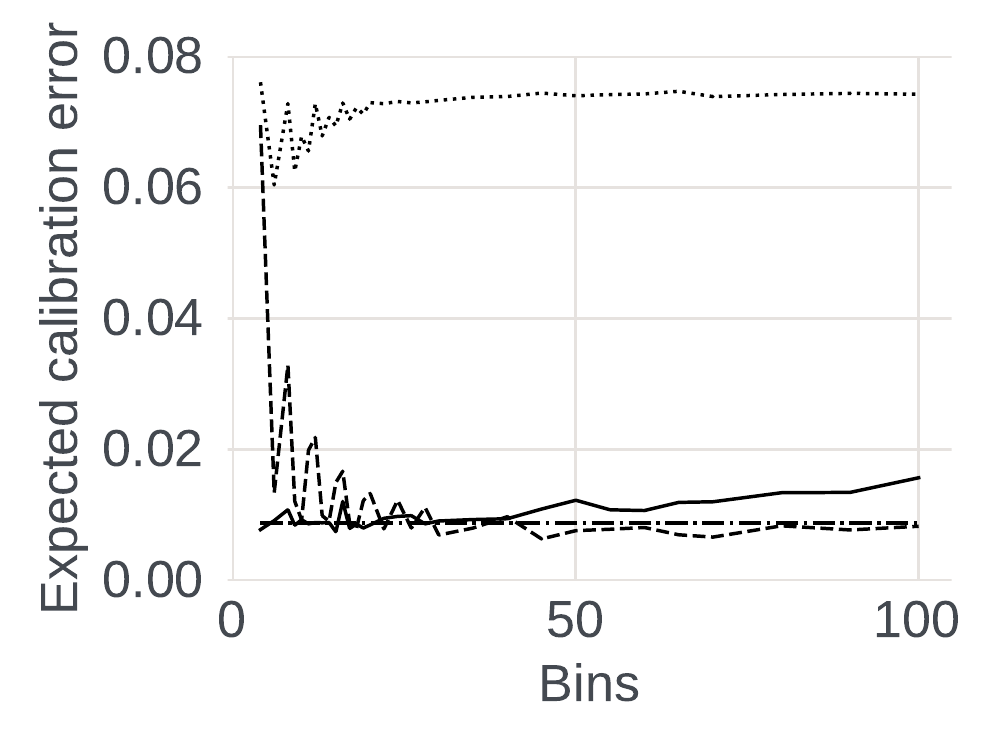}}

\hspace*{-9mm}
\subcaptionbox{Sep data, \cdp noise\label{calib-sep-lap2}}{\includegraphics[width=\figwidth]{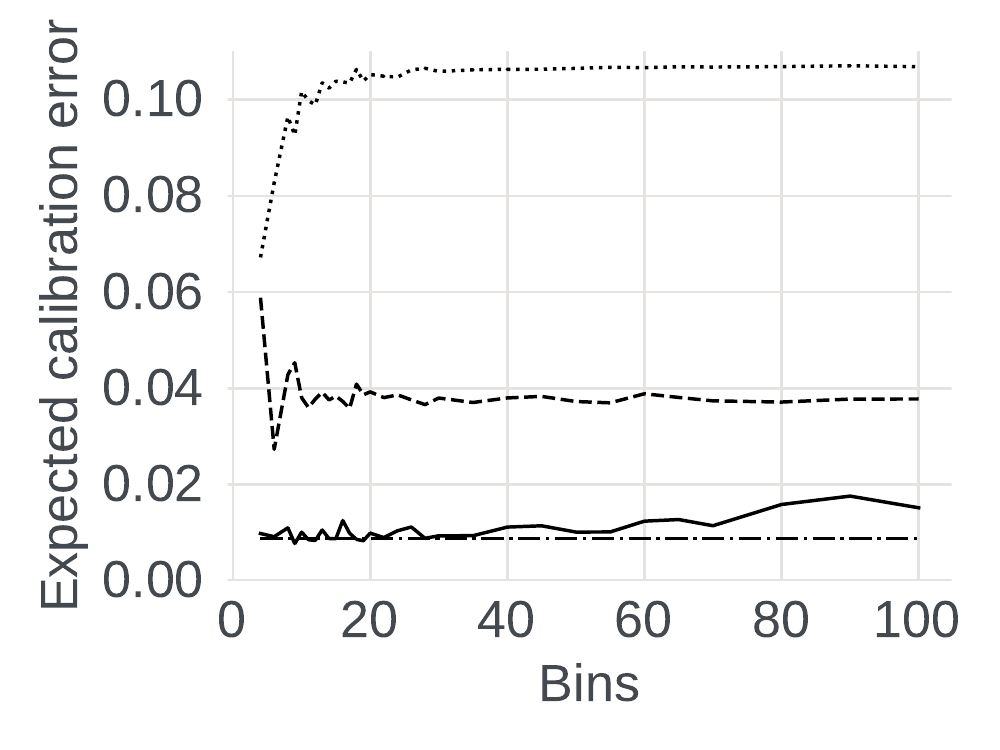}}%
\subcaptionbox{Oct data, \cdp noise\label{calib-oct-lap2}}{\includegraphics[width=\figwidth]{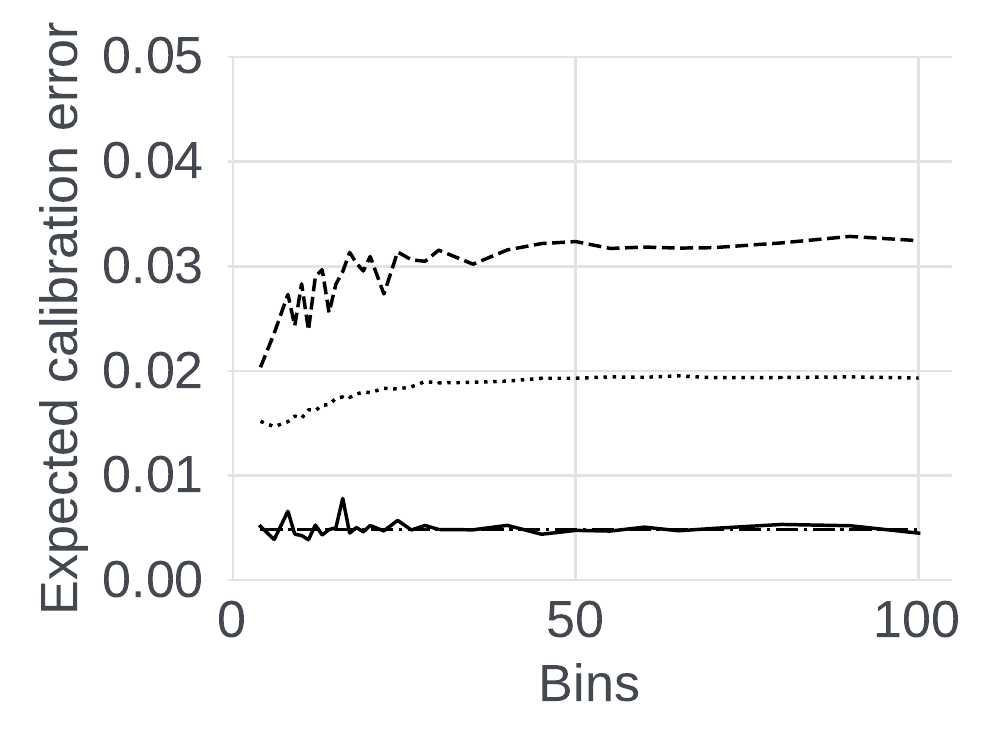}}%
\subcaptionbox{Nov data, \cdp noise\label{calib-nov-lap2}}{\includegraphics[width=\figwidth]{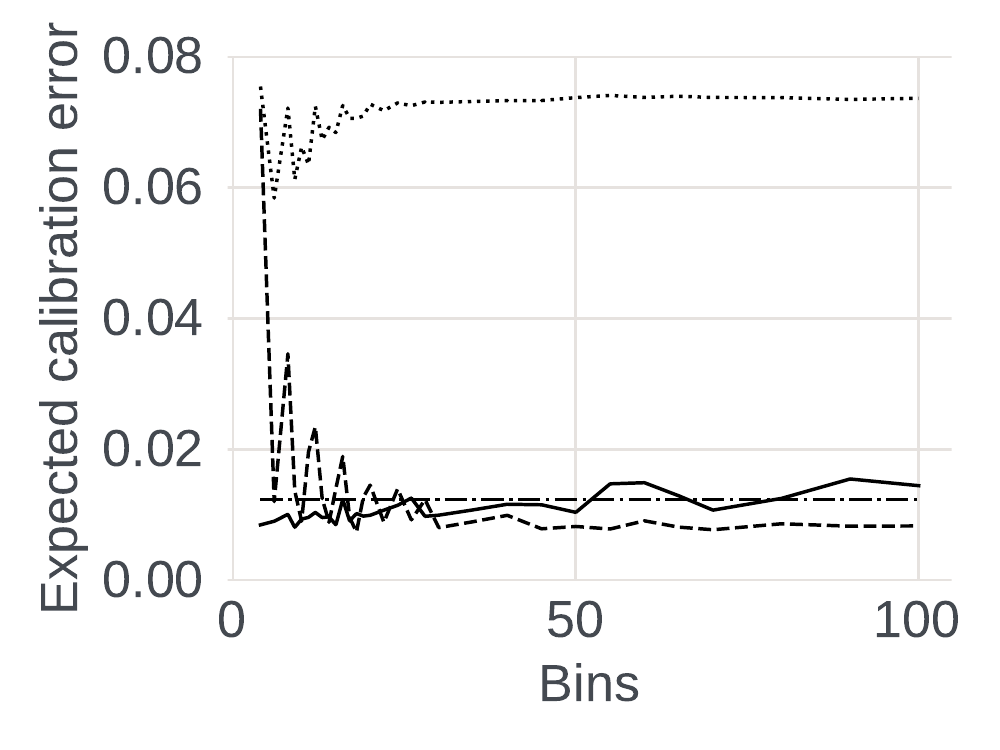}}

\hspace*{-9mm}
\subcaptionbox{Sep data, \ldp noise\label{calib-sep-oue2}}{\includegraphics[width = \figwidth]{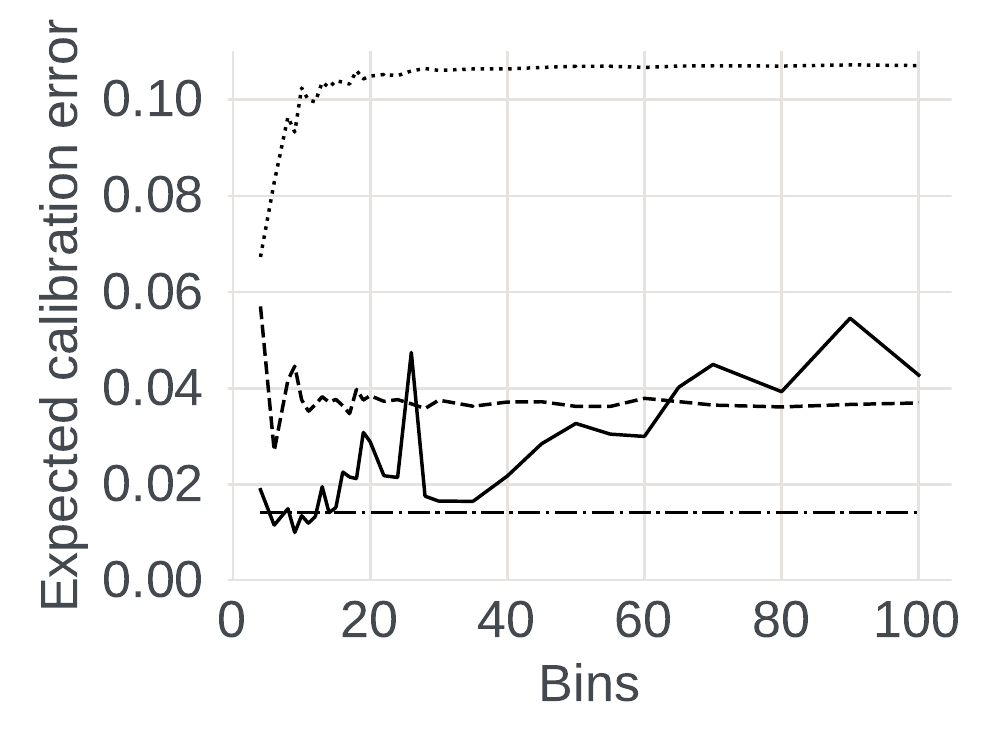}}%
\subcaptionbox{Oct data, \ldp noise\label{calib-oct-oue2}}{\includegraphics[width = \figwidth]{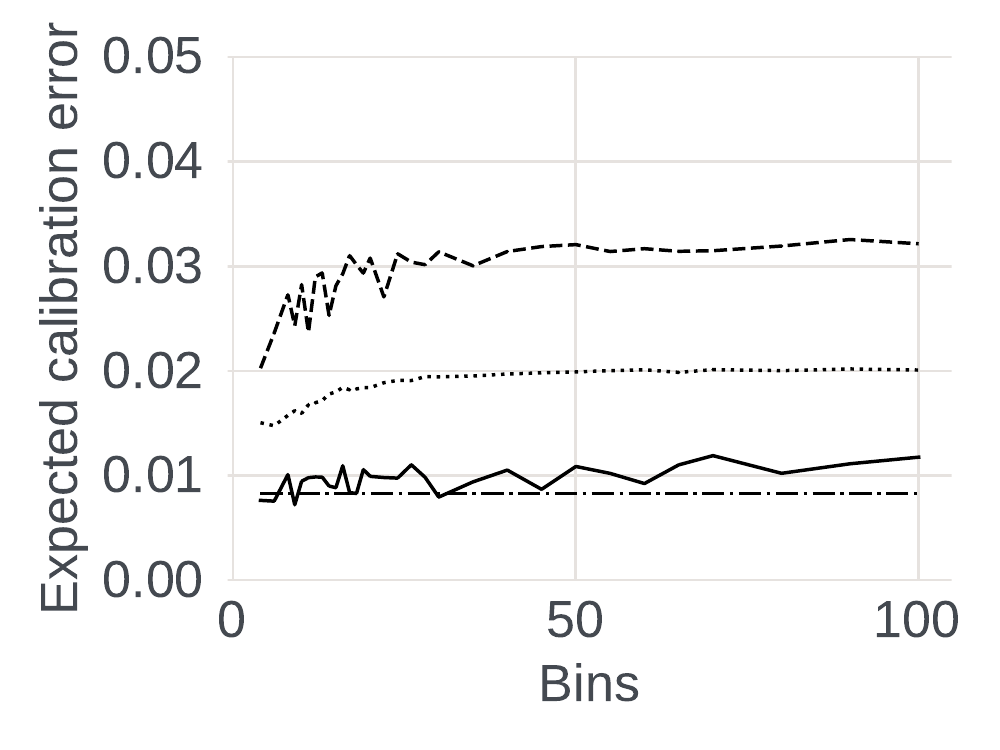}}%
\subcaptionbox{Nov data, \ldp noise\label{calib-nov-oue2}}{\includegraphics[width = \figwidth]{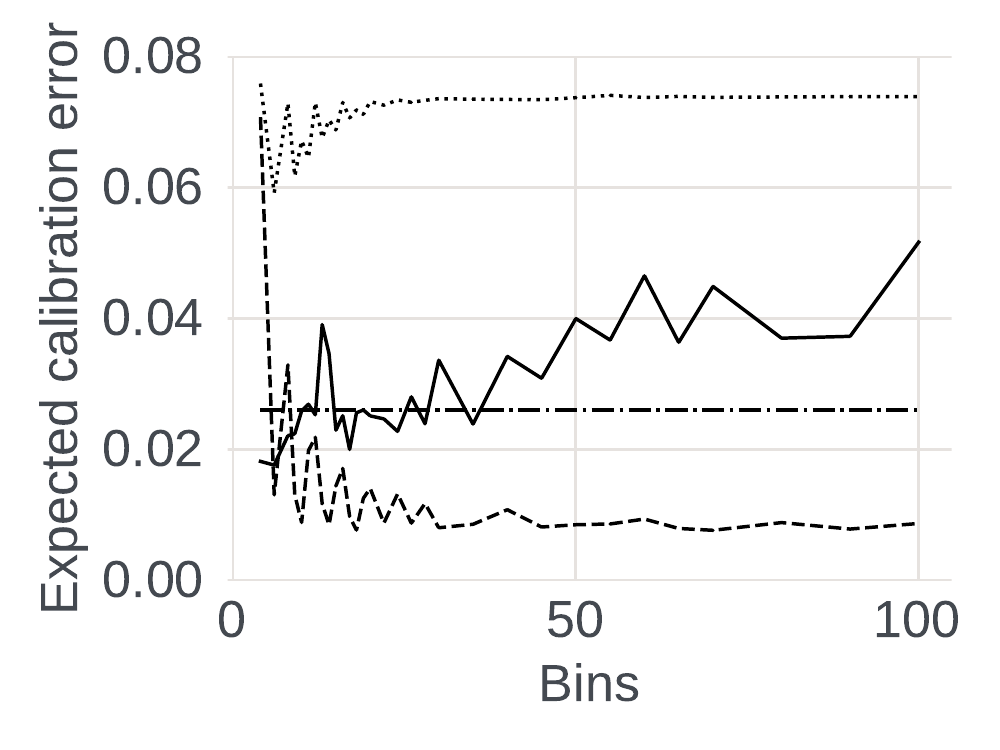}}
\caption{Classifier calibration accuracy with varying noise levels}
\label{fig:calib}
\end{figure*}

\eat{
\begin{figure*}[th]
\centering
\includegraphics[trim=800 25 800 25,clip,width=0.6\textwidth]{figs/calib_pop_legend.pdf}\newline
\subcaptionbox{Sep data\label{calib-sep-pop}}{\includegraphics[width = \figwidth]{figs/calib-sep-pop.pdf}}%
\subcaptionbox{Oct data\label{calib-oct-pop}}{\includegraphics[width = \figwidth]{figs/calib-oct-pop.pdf}}%
\subcaptionbox{Nov data\label{calib-nov-pop}}{\includegraphics[width = \figwidth]{figs/calib-nov-pop.pdf}}
\caption{Classifier calibration accuracy with varying population size}
\label{fig:calib-pop}
\end{figure*}
}

\subsection{Area Under Curve}
\label{app:aucexpts}
For Area Under Curve (AUC), we show the results over ten repetitions, varying ($i$) how the examples are sampled, and ($ii$) the random noise. For all methods, we use a hierarchy with $h=10$, found earlier to be a good choice. Figure~\ref{fig:auc} shows our results for AUC, as we vary the data and the noise model. Each plot, the guideline $1/2B$ represents the pessimistic bound from Lemma~\ref{lem:auc}, while the guideline $1/3B^2$ shows our tighter bound under the {\em well-behaved} assumption (Theorem~\ref{thm:aucbetter}). 
For each experiment, we plot a line showing the worst-case uncertainty in our estimate, due to the noise in each bucket. 
That is, the quantity corresponding to $\sum_i p_i n_i$, the sum over buckets of the product of the number of positive and negative examples.  
This is the error we would see if the analysis in Lemma~\ref{lem:auc} was tight. 
We also plot one curve for using the histogram naively, i.e., picking $B$ buckets with uniform boundaries, and the observed error for our approach where we pick $B$ buckets based on the (estimated) quantile boundaries. 
\red{This uniform choice of buckets is equivalent to the approach proposed recently by~\citet{Sunetal2} in the local model: as we will see, it is outperformed by the quantile histogram approach.}

Our first observation in the \fed case (no explicit privacy noise) is that 
the worst-case error bound indeed follows $O(1/B)$, but our tighter analysis yields errors close to $O(1/B^2)$. That is, the total uncertainty follows the $1/2B$ curve closely, while the histogram estimators follow the $1/3B^2$ curve. 
The error vanishes rapidly: with 100 buckets, the AUC is estimated with $10^{-5}$ accuracy, sufficient for most conceivable applications. 
Quantile histograms clearly outperform uniform-bucket boundaries, with up to an order of magnitude smaller error.

Our analysis in Theorem~\ref{thm:auccdp} predicts a limit to the accuracy obtainable with more buckets, due to a fixed level of noise from \cdp privacy.  
Experiments confirm this: the error initially follows the $1/3B^2$ curve, but the error curve flattens after about 20-40 buckets.  
Here, the total error in AUC estimation is $\approx$0.001 -- small enough for useful conclusions about the classifier. 
With more buckets, examples distribute across buckets without large clusters, helping the uniform histogram work as well as the quantile-based histogram.

The same behavior holds for the \ldp case, where the error bound converges to $\approx$ 0.005. The speed of convergence and value reached vary based on the data used. 
Beyond 20 buckets, the error reduction is minimal, as \ldp noise has stronger impact than the \cdp noise.  

In Figure~\ref{fig:auc-pop} we show the impact of varying $\clients$ on each of the data sets. 
Each plot shows the error obtained against ground truth for \fed, \cdp and \ldp privacy.  
We fix the number of histogram buckets to $B=100$, and use $h=10$. We see no change trend for
accuracy in the \fed case, consistent with Theorem~\ref{thm:aucbetter}. With \cdp guarantees, the error is orders of magnitude greater and decreases as $O(1/\clients)$, as predicted by Theorem~\ref{thm:auccdp}. 
Extrapolating this behavior, several million clients would be needed to match the error in the noiseless case. The pattern is similar for \ldp noise, although perhaps not as pessimistic as the $O(1/\sqrt{\clients})$ bound would suggest (Theorem~\ref{thm:aucldp}). 

\subsection{Calibration}
\label{app:calibexpts}
We seek to minimize the expected calibration error, estimated by dividing the domain of the score function into bins (distinct from the histogram buckets in our algorithms), and {comparing the observed fraction of positive examples in each bin with the average score for that bin.  This fraction is then averaged over all bins.} 
Plots in Figure~\ref{fig:calib} vary the number of such bins and allow our calibration approach to use the same number of buckets as bins. \eat{-- so if the evaluation uses 20 bins, our approach will also try to estimate the correct frequency using this many buckets.}
We can adapt the approach of \citet{NaeiniCH15}, which also makes use of histograms based on quantiles. 
The Bayesian Binning into Quantiles (BBQ) technique considers a range of choices of $B$, 
from $\clients^{1/3}/10$ up to $10\clients^{1/3}$. 
Each choice of $B$ is assigned a BBQ score based on the number of positive and negative examples in each bucket, via the Gamma function. 
The BBQ calibration function is given by a normalized sum of each binning in turn weighted by its BBQ score value.
We can blindly apply this approach in our setting, since the information gathered in the form of a high resolution histogram can be used to build the necessary score histograms for many choices of $B$. 
In the \fed case, this should give the same results as BBQ in the centralized setting. 
For histograms with privacy noise, we may expect to see some deviation in performance, since the BBQ method is not tuned to correct for the noise in bucket counts. 
On these plots, we show the calibration error identified by the Bayesian Binning into Quantiles (BBQ) method combined with our histogram approach. 
The baselines are ($i$) the calibration error of the uncalibrated score function\eat{ -- we should hope to do better than this! --} and ($ii$) the result of using the (centralized) implementation of isotonic regression from {\tt scikit-learn 0.22}. 

For the \fed case, the plots in Figures~\ref{calib-sep-nodp2}-\ref{calib-nov-nodp2} show that good accuracy is possible -- calibration error of $\approx$0.01 is achievable, i.e.,
on average, the calibrated score is within 0.01 of the true probability. 
This outcome is not very sensitive to the number of evaluation bins. 
The BBQ approach on top of our histogram approach does a good job at combining information from multiple bucketings when there is no noise, and gives a reliable choice of calibration. 
We observe that, due to the use of the Gamma function in defining scores, it is often the case that one bucketing achieves a vastly greater BBQ score than other choices. 
Then normalized weighting puts all weight on this bucketing, so the method effectively simplifies into choosing the number of buckets. 
This gives an improvement over using the uncalibrated score function, where the calibration error can be much larger, $\approx$0.1 for Sep and 0.08 for Nov. Surprisingly, the (centralized) isotonic approach is not a good fit for these score functions. 
On Sep data, it attains calibration error of 0.04, and for  Oct data it increases the error compared to the original score function. Isotonic regression clearly helps only for the Nov data.

Introducing \cdp noise does not change the results much, as anticipated by our observation in Theorem~\ref{thm:calib:dp} that privacy noise would be outweighed by the variation of data points within the bins. 
Further, the overall calibration error is similar in magnitude to the \fed case, $\approx$0.01. 

For \ldp noise, the error increases to 0.02 and higher, as the impact of privacy noise is noticeable. Given the choice of the number of buckets, using fewer calibration buckets reduces noise. Despite cruder calibration, using $\leq$10 buckets keeps the error near 0.02. 
As expected, the BBQ approach is impacted by the extra noise, and tends to end up placing more weight on choices with more buckets. 
For Oct data, the original uncalibrated score function has smaller error, and the combination of \ldp noise with calibration causes more harm than good. 
For Sep and Nov, where the original score function was not well calibrated,
federated calibration brings significant benefit.

Figure~\ref{fig:calib-pop} varies the number of clients with a fixed number of bins (20), and studies the expected calibration error. 
We see that all methods benefit from increased population size, as sampling within the bins becomes
more accurate. The relative ordering of the different privacy regimes is consistent: the least error for \fed, and the highest error for \ldp. 
The error for \cdp is not identical to \fed, but the two appear to converge for larger client populations. 
Theorem~\ref{thm:calib:dp} predicts that \ldp converges more slowly as a function of $\clients$. 

\subsection{Other experimental observations}

We briefly \red{comment on} other observations from the experimental study. 

\para{Dependence on privacy parameter $\epsilon$.} 
Table~\ref{tab:results} summarizes the bounds for \cdp and \ldp estimations as a function of varying the privacy parameter $\epsilon$. 
We found that these bounds were closely followed in our experiments when we varied $\epsilon$. 
This is unsurprising, since the impact of varying $\epsilon$ for histograms is well-understood, and the impact on accuracy is quite direct. 

\para{Time cost.}
Simulations were performed on a single CPU machine, and were not highly optimized for performance. 
Nevertheless, we accurately simulated the tasks of each client and the server within the protocol. 
Typical experiments took a matter of minutes to evaluate a large range of parameter choices and repetitions, meaning that the cost per client is trivial (\red{milliseconds of computation effort per client}), and the effort for the server is only some simple aggregation.

\end{document}